\documentclass[prd,superscriptaddress,amsfonts,amssymb,amsmath,showpacs,twocolumn]{revtex4}
\usepackage{epsfig}
\usepackage[applemac]{inputenc}
\usepackage{graphicx}

\newcommand{\pp}[1]{\left( #1 \right)}
\newcommand{\cc}[1]{\left[ #1 \right]}
\newcommand{\ch}[1]{\left\{ #1 \right\}}

\newcommand{\del}{\partial}
\newcommand{\vt}[1]{\textbf{\textrm{#1}}}
\newcommand{\ket}[1]{|{#1}\rangle}
\newcommand{\bra}[1]{\langle{#1}|}
\newcommand{\braket}[2]{\langle{#1}|{#2}\rangle}

\date{\today}

\begin{document}

\title{From quantum to classical instability in relativistic stars}

\author{Andr\'e G.\ S.\ Landulfo}\email{andre.landulfo@ufabc.edu.br}
\affiliation{Centro de Ci\^encias Naturais e Humanas,
Universidade Federal do ABC, Avenida dos Estados, 5001, CEP 09210-580, Bangu,
Santo Andr\'e, S\~ao Paulo, Brazil}
\author{William C.\ C.\ Lima}\email{wccl@ift.unesp.br}
\affiliation{Instituto de F\'\i sica Te\'orica, Universidade 
Estadual Paulista,
Rua Dr.\ Bento Teobaldo Ferraz, 271, Bloco II, CEP 01140-070, 
S\~ao Paulo, S\~ao Paulo, Brazil}
\author{George E.\ A.\ Matsas}\email{matsas@ift.unesp.br}
\affiliation{Instituto de F\'\i sica Te\'orica, Universidade 
Estadual Paulista,
Rua Dr.\ Bento Teobaldo Ferraz, 271, Bloco II, CEP 01140-070, 
S\~ao Paulo, S\~ao Paulo, Brazil}
\author{Daniel A.\ T.\ Vanzella}\email{vanzella@ifsc.usp.br}
\affiliation{Instituto de F\'\i sica de S\~ao Carlos,
Universidade de S\~ao Paulo, Caixa Postal 369, CEP 13560-970, 
S\~ao Carlos, S\~ao Paulo, Brazil}


\begin{abstract}

It has been shown that gravitational fields produced by realistic classical-matter distributions can force quantum
vacuum fluctuations of some nonminimally coupled free scalar fields to undergo a phase of exponential growth. The
consequences of this unstable phase for the background spacetime have not been addressed so far due to known
difficulties concerning backreaction in semiclassical gravity. It seems reasonable to believe, however, that the quantum
fluctuations will ``classicalize'' when they become large enough, after which backreaction can be treated in the
general-relativistic context. Here we investigate the emergence of a classical regime out of the quantum field evolution
during the unstable phase. By studying the appearance of classical correlations and loss of quantum coherence, we show
that by the time backreaction becomes important the system already behaves classically. Consequently, the
gravity-induced instability leads naturally to initial conditions for the eventual classical description of the
backreaction. Our results give support to previous analyses which treat classically the instability of scalar fields in
the spacetime of relativistic stars, regardless of whether the instability is triggered by classical or quantum
perturbations.
 
\end{abstract}

\maketitle

\section{Introduction}

The vacuum state of quantum fields harbors many interesting physical phenomena. In particular, the vacuum gravitates,
meaning, according to general relativity, that it affects and is affected by the spacetime geometry. Although this fact
leads to important consequences for cosmology and black hole physics, it normally produces only feeble effects at
astrophysical scales. In contrast to this, it was argued in Ref.~\cite{lv_prl_2010} that well behaved spacetimes curved
by classical matter may induce vacuum fluctuations of some nonminimally coupled free scalar fields to go through a phase
of exponential growth. This growth enhances the expectation value of the field energy-momentum tensor, eventually
leading the vacuum to take over the system evolution. A concrete realization of this claim was given in
Ref.~\cite{lmv_prl_2010}, where the amplification of the vacuum fluctuations was studied in the spacetime of a
relativistic star. The appearance of this instability in other astrophysically inspired scenarios was explored in
Refs.~\cite{lmmv_prd_2013,mmv_2014b}. As the system is driven to a new equilibrium state, a burst of free scalar
particles is expected, regardless of the details of the final configuration \cite{llmv_prd_2012}. Nevertheless, the
final configuration is important for astrophysical purposes. In order to determine it, one must take into account the
backreaction of the quantum field on the spacetime. This is a highly nontrivial task due to the well known difficulties
concerning the backreaction in semiclassical gravity.

Notwithstanding, it seems reasonable to believe that quantum fluctuations amplified enough to menace the stability
of relativistic stars cannot remain ``quantum'' for too long. Thus, if the quantum phase ends before vacuum fluctuations
dominate the system, we expect backreaction to be well described by the classical general-relativistic equations. With
this scenario in mind, we investigate the quantum-to-classical transition of the quantum fluctuations in the vacuum
state, showing that the system does classicalize prior semiclassical backreaction becomes paramount. 

The transition of a quantum system to a regime in which its behavior is well approximated by classical physics is a
matter that has received attention in different areas --- see, e.g., Ref.~\cite{gjkksz_book_1996}. For this
quantum-to-classical transition to happen, two ingredients, normally related, are necessary: the appearance of certain
classical correlations and the loss of quantum coherence. By classical correlations we mean that the corresponding
Wigner function is peaked at the classical trajectories, while the loss of quantum coherence is necessary to forbid
their superposition. The loss of quantum coherence, in particular, results from the entanglement of the system with
other ``unobservable'' degrees of freedom which are eventually traced out. Thus, in order to study the emergence of a
given classical behavior from a quantum system it is unavoidable to take into account its interaction with some set of
additional degrees of freedom, generally referred to as ``environment'' \cite{z_pht_1991}. Both the appearance of
classical correlations and the process of decoherence in this open system depend, in principle, on the system internal
dynamics and its interaction with the environment. The form of the interaction is particularly important: it defines
according to what observable the system will be regarded as classical \cite{z_prd_1981_1982}. 

Here, we are interested in the behavior of the unstable modes of the scalar field, since they dominate the vacuum
fluctuations. We will show that for these modes the internal dynamics will be enough to produce classical correlations.
The decoherence process needed to ensure the quantum-to-classical transition, as defined above, will depend on the
interaction of the scalar field with gravity. The most natural environment to consider is the one formed by the quantum
fluctuations of the background metric --- gravitons. These will not be the only degrees of freedom of our environment,
though. The coupling of the scalar field with gravity induces an interaction between the unstable and stable modes of
the scalar field, making the latter ones also part of the environment. From this analysis we can estimate the time scale
for the unstable modes to become classical with respect to their amplitude and canonically conjugate momentum. This
time scale is of fundamental importance to determine whether backreaction may be treated in the classical rather than
semiclassical realm. 

The paper is organized as follows. In Sec.\ \ref{sec:gravity_induced_instability} we briefly revisit the quantization of
an unstable free scalar field nonminimally coupled to gravity in the spacetime of a relativistic star. In Sec.\
\ref{sec:free_evolution} we focus on the sector of the Fock space related to the unstable modes and describe the
evolution of the corresponding vacuum state through its Wigner function representation. It is shown that the field
amplitude and its canonically conjugate momentum become classically correlated in a time scale comparable to the one set
by when backreaction becomes important. Next, in Sec.\ \ref{sec:decoherence} we discuss the loss of coherence of the
vacuum fluctuations. By integrating out the degrees of freedom of the gravitons and of the stable modes of the scalar
field, we obtain a master equation for the density matrix describing the state of the unstable modes. The analysis of
this master equation shows that by the time backreaction becomes important the initially pure vacuum state has already
evolved into a mixture of localized states in field amplitude and momentum. We close the discussion and make our final
remarks in Sec.\ \ref{sec:final_remarks}. Throughout the text we shall assume that $\hbar=c=1$, and the signature
$(-+++)$ for the spacetime metric.

\section{Gravity-induced instability}
\label{sec:gravity_induced_instability}

We start by considering a real scalar field $\phi$ evolving over a globally hyperbolic spacetime background
$(\mathcal{M},g_{ab})$ curved by some classical-matter distribution. The field obeys the Klein-Gordon equation,
\begin{equation}\label{kg_eq}
 -\nabla_a\nabla^a\phi+(m^2+\xi R)\phi=0,
\end{equation}
where $m\ge0$ is the field mass, $\xi\in\mathbb{R}$ is the nonminimal coupling parameter, and $R$ stands for the
scalar curvature. The associated energy-momentum tensor is given by
\begin{eqnarray}\label{scalar_field_emt}
 T_{ab}&=&(1-2\xi)\nabla_a\phi\nabla_b\phi+\xi R_{ab}\phi^2-2\xi \phi\nabla_a\nabla_b\phi\nonumber\\
&&+\pp{2\xi-\frac{1}{2}}g_{ab}[\nabla_c\phi\nabla^c\phi+(m^2+\xi R)\phi^2],
\end{eqnarray}
where $R_{ab}$ stands for the Ricci tensor.

We quantize the field $\phi$ according to the canonical procedure. Then, the field operator $\hat{\phi}$ can be expanded
in terms of a complete set of positive- and negative-norm solutions
$\{u_{\alpha}^{(+)},u_{\alpha}^{(-)}\}_{\alpha\in\mathcal{I}}$, 
\begin{equation}\label{quantum_field}
 \hat{\phi}=\int{d\vartheta(\alpha)[\hat{a}_\alpha u_\alpha^{(+)}+\hat{a}_\alpha\phantom{}^\dagger
u_\alpha^{(-)}]},
\end{equation}
with $u_\alpha^{(+)}$ and $u_\alpha^{(-)}\equiv u_\alpha^{(+)*}$ orthonormalized according to the Klein-Gordon inner
product. Here, $\mathcal{I}$ stands for some set of good quantum numbers, while $\vartheta$ denotes some measure over
this set. As usual, the canonical commutation relations combined with the completeness of the modes imply that the
creation and annihilation operators $\hat{a}_\alpha\phantom{}^\dagger$ and $\hat{a}_\alpha\phantom{}$, respectively,
satisfy
\begin{equation}\label{ccr}
 [\hat{a}_\alpha,\hat{a}_\beta\phantom{}^\dagger]=\delta_{\vartheta}(\alpha,\beta),
\end{equation}
while other commutators vanish. The $\delta_{\vartheta}$ denotes the delta distribution according to the measure
$\vartheta$, i.e., $\int{d\vartheta(\alpha)\mathfrak{f}(\alpha)\delta_{\vartheta}(\alpha,\beta)}=\mathfrak{f}(\beta)$.
Finally, the vacuum state associated with the selected set of modes is defined by demanding $\hat{a}_\alpha\ket{0}=0$
for all $\alpha\in\mathcal{I}$. 

Assuming that the background spacetime is curved by the presence of a static, spherically symmetric compact object,
its metric can be written as
\begin{equation}\label{star_metric}
 ds^2=-f(dt^2-d\chi^2)+r^2(d\theta^2+\sin^2\theta d\varphi^2),
\end{equation}
where $f=f(\chi)>0$ and $r=r(\chi)\ge 0$ are functions of the radial coordinate $\chi$ such that
$\lim_{\chi\rightarrow+\infty}f(\chi)=1$, $\lim_{\chi\rightarrow+\infty}r(\chi)/\chi=1$, and $dr/d\chi>0$. The last
requirement prevents the existence of trapped surfaces. By using the symmetries of the underlying spacetime, it is
possible to find a set of time-oscillating positive-norm solutions of Eq.~(\ref{kg_eq}) with the form
\begin{equation}\label{v_modes}
 v_{\varpi l\mu}^{(+)}(t,\vt{x})=\frac{e^{-i\varpi t}}{\sqrt{2\varpi}}\frac{\psi_{\varpi l}(\chi)}{r(\chi)}
Y_{l \mu}(\theta,\varphi),
\end{equation}
where $\vt{x}$ denotes the spatial coordinates, while $Y_{l \mu}$ stands for the spherical harmonics, with
$l=0,1,2,\dots$ and $\mu=-l,-l+1,\dots,l$, and $\varpi>0$. The radial part of $v_{\varpi l\mu}^{(+)}$ satisfies
\begin{equation}\label{psi_varpi_eq}
 -\frac{d^2}{d\chi^2}\psi_{\varpi l}+V^{(l)}_{\textrm{eff}} \psi_{\varpi l}=\varpi^2 \psi_{\varpi l},
\end{equation}
vanishing at the origin and being well behaved at spatial infinity. For a star composed of perfect fluid, one can use
Einstein equations to cast the effective potential $V^{(l)}_{\textrm{eff}}$ in Eq.~(\ref{psi_varpi_eq}) as
\begin{equation}\label{V_eff}
 V^{(l)}_{\textrm{eff}}=f\cc{m^2+\frac{l(l+1)}{r^2}+\pp{\xi-\frac{1}{6}}R+\frac{8\pi G}{3}(\bar{\rho}-\rho)},
\end{equation}
where $\rho=\rho(\chi)$ denotes the energy density of the stellar fluid and
\begin{equation}\label{av_den}
 \bar{\rho}(\chi)\equiv\frac{3M(\chi)}{4\pi r^3(\chi)}
\end{equation}
is the average density of the star up to the radial coordinate $r(\chi)$, which encompasses a mass $M(\chi)$. 

Depending on (i) the values of the nonminimal coupling parameter $\xi$, (ii) the mass-radius ratio of the star, and
(iii) its equation of state, the time-oscillating modes (\ref{v_modes}) may not be the only ones complying with the
boundary conditions mentioned above. Indeed, the effective potential (\ref{V_eff}) allows the existence of ``bound
states'' \cite{lmv_prl_2010}. These solutions give rise to exponentially growing modes, 
\begin{equation}\label{w_modes}
 w_{\Omega l\mu}^{(+)}(t,\vt{x})=\frac{e^{\Omega t-i\pi/4}+e^{-\Omega t+i\pi/4}}{\sqrt{4\Omega}}
\frac{\psi_{\Omega l}(\chi)}{r(\chi)}Y_{l\mu}(\theta,\varphi),
\end{equation}
for which the radial part obeys
\begin{equation}\label{psi_Omega_eq}
 -\frac{d^2}{d\chi^2}\psi_{\Omega l}+V^{(l)}_{\textrm{eff}} \psi_{\Omega l}=-\Omega^2 \psi_{\Omega l},
\end{equation}
with $\Omega>0$ and the form of the temporal part was chosen to ensure the positivity of the $w_{\Omega l\mu}^{(+)}$
norm \cite{footnote_1}. 

For the sake of simplicity, we shall assume hereafter the existence of a single unstable mode. Since the centrifugal
term in Eq.~(\ref{V_eff}) contributes positively to the effective potential, this mode will have angular momentum
quantum numbers $l=\mu=0$ and will be denoted simply by $w_{\Omega}^{(+)}$ and its radial part by $\psi_{\Omega}/r$. The
spatial part of the stable and unstable modes will be denoted by $F_{\varpi l\mu}$ and $F_\Omega$, respectively, i.e.,
$F_{\varpi l\mu}(\vt{x})\equiv \psi_{\varpi l} Y_{l\mu}/r$ and $F_\Omega(\vt{x})\equiv \psi_\Omega Y_{00}/r$. We shall
denote by $\hat{a}_{\varpi l\mu}\phantom{}^\dagger$ and $\hat{a}_{\varpi l\mu}$ the creation and annihilation operators
defined by the modes $v_{\varpi l\mu}^{(+)}$ and by $\hat{a}_\Omega\phantom{}^\dagger$ and $\hat{a}_\Omega$ the same
operators defined by the mode $w_\Omega^{(+)}$.

Let us consider the situation in which the system begins in a stationary stable phase in the past and evolves into an
unstable one in a time scale much smaller than any other present in the problem. Assuming that the quantum field is in
the vacuum state with respect to the stationary past observers, it is possible to show that the initially quiescent
quantum vacuum fluctuations will grow as $\langle\hat{\phi}^2\rangle\propto e^{2\Omega t}$ during the unstable phase.
The exponential enhancement of the quantum fluctuations impacts on the (renormalized) expectation value of the
energy-momentum tensor operator, $\langle \hat{T}_{ab}\rangle$, eventually leading the quantum field to backreact on the
spacetime \cite{lv_prl_2010}. In order to estimate how long it takes for the quantum fluctuations to threaten the star
stability, we first note that the existence of the unstable (bound) solutions typically requires potentials satisfying
$\sup|V_{\textrm{eff}}^{(0)}|\mathsf{R}^2\sim1$, in which case $\Omega^2\sim \sup|V_{\textrm{eff}}^{(0)}|\sim
\mathsf{R}^{-2}$, where $\mathsf{R}$ denotes the radial coordinate $r$ of the star surface --- see discussion in
Sec.~III of Ref.~\cite{llmv_prd_2012}. By calculating, e.g., the ratio between the vacuum and stellar energy densities, 
\begin{equation}\label{vacuum_star_ratio}
 \frac{\rho_{\textrm{v}}}{\rho}\sim\pp{\frac{\ell_{\textrm{P}}}{\mathsf{R}}}^2\times
\exp{(2t/\mathsf{R})},
\end{equation}
with $\ell_\textrm{P}$ denoting the Planck length, one concludes that the backreaction time scale dictated by the
semiclassical Einstein equations is $t_{\textrm{br}}\sim \mathsf{R}\ln(\mathsf{R}/\ell_\textrm{P})$, which is of
the order of a few milliseconds for a neutron star \cite{lmv_prl_2010} --- for a more comprehensive account on this
vacuum awakening effect, see Ref.~\cite{llmv_prd_2012}, and Refs.~\cite{l_prd_2013,ss_prd_1970} for a rigorous
discussion on the quantization of unstable linear fields in globally hyperbolic spacetimes.

\section{Free field evolution and the appearance of classical correlations}
\label{sec:free_evolution}

In a static spacetime, like the one engendered by the relativistic star considered above, the Hamiltonian operator can
be formally defined from the energy-momentum tensor as
\begin{equation}\label{hamiltonian_operator_definition}
 \hat{H}\equiv\int_{\Sigma}{d\Sigma n^a\varkappa^b \hat{T}_{ab}}.
\end{equation}
Here, $\varkappa^a=(\del_t)^a$ is the Killing vector field generating the time isometry, $n^a$ is a future-pointing
unit vector field orthogonal to the Cauchy surface $\Sigma$, and $d\Sigma\equiv \sqrt{h}d^3x$ is the volume
element with respect to the spatial metric tensor $h_{ab}$, with $h\equiv\textrm{det}\, h_{ab}$. Using the expansion
(\ref{quantum_field}) in terms of the modes $v_{\varpi l\mu}^{(+)}$ and $w_\Omega^{(+)}$ , we obtain from
Eq.~(\ref{hamiltonian_operator_definition}) that
\begin{equation}\label{hamiltonian_operator}
 \hat{H}=\hat{H}_\textrm{s}+\hat{H}_\textrm{u}.
\end{equation}
In Eq.~(\ref{hamiltonian_operator}) the Hamiltonian operator associated to the unstable mode $w_\Omega^{(+)}$ is given
by
\begin{equation}\label{h_u}
\hat{H}_\textrm{u}=-\frac{\Omega}{2}(\hat{a}_\Omega\hat{a}_\Omega
+\hat{a}_\Omega\phantom{}^\dagger\hat{a}_\Omega\phantom{}^\dagger),
\end{equation} 
which corresponds to the Hamiltonian of an upside-down harmonic oscillator, while $\hat{H}_\textrm{s}$ is the
Hamiltonian operator related to the stable modes $v_{\varpi l\mu}^{(+)}$ and consists of a collection of harmonic
oscillators. Hence, we shall revisit the quantum upside-down harmonic oscillator in the light of our problem.

By defining the operators
\begin{equation}\label{q_definition}
 \hat{q}_\Omega\equiv\frac{1}{\sqrt{2\Omega}}(\hat{a}_{\Omega}+\hat{a}_{\Omega}\phantom{}^\dagger)
\end{equation}
and
\begin{equation}\label{p_definition}
 \hat{p}_\Omega\equiv-i\sqrt{\frac{\Omega}{2}}(\hat{a}_{\Omega}-\hat{a}_{\Omega}\phantom{}^\dagger),
\end{equation}
Eq.~(\ref{h_u}) can be cast as
\begin{equation*}
 \hat{H}_\textrm{u}=\frac{1}{2}\hat{p}_\Omega^2-\frac{\Omega^2}{2}\hat{q}_\Omega^2.
\end{equation*}
The operators (\ref{q_definition}) and (\ref{p_definition}) are related to the field operator $\hat{\phi}$ and its time
derivative according to
\begin{equation*}
 \hat{q}_\Omega=\frac{1}{\sqrt{4\pi}}\int_{0}^{+\infty}rd\chi\int_{0}^{\pi}\sin\theta
d\theta \int_{0}^{2\pi}d\varphi\hat{\phi}(0,\vt{x})\psi_\Omega(\chi)
\end{equation*}
and
\begin{equation*}
 \hat{p}_\Omega=\frac{1}{\sqrt{4\pi}}\int_{0}^{+\infty}rd\chi\int_{0}^{\pi}\sin\theta
d\theta \int_{0}^{2\pi}d\varphi\del_t\hat{\phi}(0,\vt{x})\psi_\Omega(\chi).
\end{equation*}
In order to obtain these expressions, we have used that $\psi_{\Omega}$ can be chosen to be a real function satisfying
\begin{equation*}
 \int_{0}^{+\infty}{d\chi \psi_\Omega(\chi)^2}=1
\end{equation*}
and
\begin{equation*}
 \int_{0}^{+\infty}{d\chi \psi_\Omega(\chi) \psi_{\varpi 00}(\chi)}=0,
\end{equation*}
for all $\varpi>0$, and the orthogonality between the spherical harmonics. It will be with respect to the observables
$\hat{p}_\Omega$ and $\hat{q}_\Omega$ that we shall investigate the classicalization of the unstable mode.

Next, we note that in this fixed-background regime the modes are decoupled and evolve independently. Thus, it is
possible to write the vacuum state of the unstable quantum field as the following tensor product: 
\begin{equation}\label{vacuum_decomposition}
\ket{0}=\ket{0_\textrm{s}}\otimes\ket{0_\textrm{u}},
\end{equation}
where $\ket{0_\textrm{s}}$ and $\ket{0_\textrm{u}}$ are defined by $\hat{a}_{\varpi l\mu}\ket{0_\textrm{s}}=0$, for all
$\varpi$, $l$ and $\mu$, and $\hat{a}_\Omega\ket{0_\textrm{u}}=0$. Therefore, one can separate the Fock space in its
stable and unstable sectors and study their time evolution separately --- see, e.g., Ref.~\cite{l_prd_2013}. 

In what follows we will focus the discussion on the evolution of the state $\ket{0_\textrm{u}}$. For this end, let us
define $\ket{\eta(t)}\equiv \hat{U}(t)\ket{0_\textrm{u}}$, with the evolution operator
\begin{equation}\label{evolution_operator}
 \hat{U}(t)\equiv e^{-it\hat{H}_\textrm{u}}.
\end{equation}
Thus, the fact that $\hat{a}_\Omega\ket{0_\textrm{u}}=0$ implies 
\begin{equation*}
\hat{U}(t)\hat{a}_\Omega\hat{U}^\dagger(t)\ket{\eta(t)}=0.
\end{equation*} 
Then, by using the identity
\begin{equation*}\label{squeezed_a}
 \hat{U}(t)\hat{a}_\Omega \hat{U}^\dagger(t)= \hat{a}_\Omega\cosh\Omega t
-i\hat{a}_\Omega\phantom{}^\dagger\sinh\Omega t
\end{equation*}
and the definitions given in Eqs.~(\ref{q_definition}) and (\ref{p_definition}), one arrives at
\begin{equation}\label{squeezed_vacuum_eq}
 \hat{p}_{\Omega}\ket{\eta(t)}=\alpha(t)\hat{q}_\Omega\ket{\eta(t)},
\end{equation}
where the function $\alpha(t)$ is conveniently written as $\alpha(t)=i[a(t)+ib(t)]$ with
\begin{equation}\label{a_funct}
 a(t)\equiv\frac{\Omega}{\cosh2\Omega t}
\end{equation}
and
\begin{equation}\label{b_funct}
 b(t)\equiv -\Omega\tanh2\Omega t.
\end{equation}
Finally, by solving Eq.~(\ref{squeezed_vacuum_eq}) in the representation of the eigenstates of $\hat{q}_\Omega$, i.e.,
solving for $\eta(t,q)\equiv\ \braket{q}{\eta(t)}$, one has
\begin{equation}\label{squeezed_vacuum}
 \eta(t,q)=\pp{\frac{a(t)}{\pi}}^{1/4}e^{i\alpha(t)q^2/2}.
\end{equation}
The wave function (\ref{squeezed_vacuum}) is known in the literature as the squeezed vacuum state. The evolution
operator defined in Eq.~(\ref{evolution_operator}) is the squeeze operator $\hat{S}(s,\beta)\equiv
\exp{\{-s(e^{-i2\beta}\hat{a}_\Omega\hat{a}_\Omega-e^{i2\beta}\hat{a}_\Omega\phantom{}^\dagger\hat{a}
_\Omega\phantom{}^\dagger)/2\}}$, with the squeezing parameter $s=\Omega t$ and the squeezing angle
$\beta=-\pi/4$. For a detailed account on the properties of these states, see, e.g.,
Refs.~\cite{zfg_rmp_1990,d_job_2002}.

A useful tool to analyze the classicalization of a quantum system is the Wigner function. Given a
general state represented by the density matrix $\hat{\varrho}$, the associated Wigner function is defined by
\cite{w_pr_1932}
\begin{equation}\label{wf_definition}
 W(t,q,p)\equiv \frac{1}{2\pi}\int_{-\infty}^{+\infty}{dy\varrho(t,q-y/2,q+y/2)e^{ipy}},
\end{equation}
with $\varrho(t,q,q^\prime)\equiv \bra{q}\hat{\varrho}(t)\ket{q^\prime}$. From Eq.~(\ref{wf_definition}), one sees that
$W$ is a real function, but not necessarily positive. Moreover, if $\hat{\varrho}=\ket{\psi}\bra{\psi}$, then
\begin{equation*}
 \int_{-\infty}^{+\infty}{dpW(q,p)}=|\psi(q)|^2
\end{equation*}
and
\begin{equation*}
\int_{-\infty}^{+\infty}{dqW(q,p)}=|\tilde{\psi}(p)|^2,
\end{equation*}
with $\tilde{\psi}(p)$ being the Fourier transform of $\psi(q)\equiv\langle q|\psi\rangle$. 

In the case of an upside-down harmonic oscillator, $W$ obeys
\begin{equation}\label{w_eq}
 \del_t W(t,q,p)=\{H(q,p),W(t,q,p)\},
\end{equation}
where $\{\ \cdot\ ,\ \cdot \ \}$ denotes the Poisson bracket and $H(q,p)$ is the classical Hamiltonian
\begin{equation}\label{class_hamiltonian}
 H(q,p)\equiv \frac{1}{2}p^2-\frac{\Omega^2}{2}q^2.
\end{equation}
We remark that Eq.~(\ref{w_eq}) coincides with the Liouville equation for the classical system defined by the
Hamiltonian (\ref{class_hamiltonian}). (We emphasize that this result holds only for quadratic potentials.) In order to
analyze Eq.~(\ref{w_eq}), it is useful to rewrite it in terms of the variables 
\begin{equation}\label{uv_variables}
u\equiv\frac{p-\Omega q}{2}\quad \textrm{and}\quad
v\equiv\frac{p+\Omega q}{2},
\end{equation}
which leads to
\begin{equation}\label{wf_uv_eq}
 \del_t W=\Omega(u\del_u-v\del_v)W.
\end{equation}
For localized states, Eq.~(\ref{wf_uv_eq}) tends to exponentially stretch the Wigner function along the $v$ direction
and exponentially squeeze it along the $u$ direction. This behavior just reflects the structure of the classical phase
space of the upside-down harmonic oscillator. The orbits of the Hamiltonian (\ref{class_hamiltonian}) are hyperbolas
with asymptotes at the lines $u=0$ and $v=0$, while $u=v=0$ is a saddle point. Physically, this means that the particles
are generally pushed away from the origin by the potential.

In our particular case, we have from Eq.~(\ref{squeezed_vacuum}) that the density matrix of the squeezed vacuum is
\begin{equation}\label{dm_squeezed_vacuum}
 \varrho(t,q,q^\prime)=\sqrt{\frac{a(t)}{\pi}}e^{-\frac{1}{2}a(t)(q^2+q^{\prime2})-\frac{i}{2}b(t)(q^2-q^{\prime2})}
\end{equation}
and, thus, the corresponding Wigner function gives
\begin{equation}\label{wf_squeezed_vacuum}
 W(t,q,p)=\frac{1}{\pi}\exp\ch{-a(t)q^2-\frac{[p+b(t)q]^2}{a(t)}}.
\end{equation} 
In the limit $\Omega t\gg1$, one has from Eqs.~(\ref{a_funct}) and (\ref{b_funct}) that $a\approx2\Omega e^{-2\Omega
t}$ and $b\approx-\Omega$. Hence, the state $\ket{0_\textrm{u}}$ evolves into a highly squeezed state and
Eq.~(\ref{wf_squeezed_vacuum}) reduces to
\begin{equation}\label{wf_squeezed_vacuum_assymp}
 W(t,q,p)\stackrel{\Omega t\gg1}{\approx}|\eta(t,q)|^2\delta(p-\Omega q),
\end{equation}
where $\delta$ denotes the usual delta distribution. As time goes by, the Wigner function (\ref{wf_squeezed_vacuum})
becomes negligibly small away from the classical trajectory $u=0$. This shows that when $\Omega
t\gg1$, the possible values for the amplitude (proportional to $q$) and momentum (proportional to $p$) of the unstable
mode are correlated along a classical trajectory in the phase space.

The expression for $W$ given in Eq.~(\ref{wf_squeezed_vacuum}) is positive, a fact that holds for any Wigner function
associated with a Gaussian state. Moreover, from Eq.~(\ref{wf_definition}), $W$ also satisfies
$\int_{-\infty}^{+\infty}{dqdpW}=1$. Thus, it can be seen as a probability distribution over the classical phase space
of the system. This interpretation, combined with the appearance of classical correlations, is sometimes regarded as a
kind of quantum-to-classical transition. In cosmology, for instance, it can account for some of the features of the
cosmic microwave background inhomogeneities \cite{gp_prd_1985,afjp_prd_1994,ps_cqg_1996}, which can be traced back to
the quantum fluctuations present in the inflationary epoch \cite{l_book_1990,pt_book_2009}. 

Notwithstanding, there are some reasons why one should regard this kind of quantum-to-classical transition as being
incomplete. For instance, looking at Eq.~(\ref{dm_squeezed_vacuum}) one sees that while $W$ becomes peaked at a
classical trajectory, the pure state $\hat{\varrho}$ turns more delocalized in both $q$ and $p$ representations.
Besides, $W$ cannot be interpreted as a probability distribution in general. (As remarked above, the Wigner function can
assume negative values, a fact directly related to interference.) Fortunately, these difficulties can be overcome if one
takes into account decoherence effects. In inflationary cosmology, a more comprehensive understanding of the
quantum-to-classical transition including decoherence was tackled, e.g., in
Refs.~\cite{kps_ijmpd_1998,klps_cqg_1998,ln_prd_2005,klps_cqg_2007,m_cqg_2007,kp_asl_2009}.

\section{Decoherence and the emergence of classical initial conditions}
\label{sec:decoherence}

In order to complete the picture of the quantum-to-classical transition, we shall analyze the loss of quantum coherence
by the scalar field during the unstable phase. As anticipated, we shall study the decoherence process induced by the
interaction between the quantum scalar field and quantum fluctuations of the gravitational field. To do so, we will
apply standard perturbative quantum field theory techniques to gravity from the perspective of an effective field theory
\cite{d_prd_1994,apv_prd_2004,k_book_2007}.

\subsection{Environment}
\label{subsec:environment}

We start by considering the classical action
\begin{equation}\label{total_action}
S=S_{\textrm{EH}}+S_{\Phi}+S_{\textrm{M}}.
\end{equation}
The Einstein-Hilbert action is given by
\begin{equation*}
 S_{\textrm{EH}}[g_{ab}]\equiv\frac{2}{\kappa^2}\int_{\mathcal{M}}{d^4x\sqrt{-g}R},
\end{equation*}
with $\kappa\equiv\sqrt{32\pi G}$ and $g\equiv\textrm{det}\, g_{ab}$, the scalar field action is defined through
\begin{equation*}
 S_{\Phi}[\Phi,g_{ab}]\equiv-\frac{1}{2}\int_{\mathcal{M}}{d^4x\sqrt{-g}[\nabla_a\Phi\nabla^a\Phi+(m^2+\xi R)\Phi^2]},
\end{equation*}
and $S_{\textrm{M}}[\Psi,g_{ab}]$ stands for the classical-matter action. We perturb this system by taking
$g_{ab}\rightarrow g_{ab}+\kappa \gamma_{ab}$ and $\Phi\rightarrow\Phi+\phi$, while keeping the classical matter
unperturbed, and expand the total action (\ref{total_action}) up to second order in both $\gamma_{ab}$ and $\phi$. In
what follows it will be assumed $\Phi=0$. (Thus, $\phi$ is small in the sense that it can only induce small
perturbations on the background metric.)

The expansion of the Einstein-Hilbert action gives
\begin{eqnarray*}
 S_{\textrm{EH}}[g_{ab}+\kappa\gamma_{ab}]&=&
\int_{\mathcal{M}}d^4x\sqrt{-g}\left[\frac{2}{\kappa^2}R+\mathcal{L}_{\textrm{EH}}^{(1)}\right.\\
&&\left.+\mathcal{L}_{\textrm{EH}}^{(2)}+\dots\right],
\end{eqnarray*}
with
\begin{equation*}
\mathcal{L}_{\textrm{EH}}^{(1)}\equiv-\frac{2}{\kappa}\pp{R_{ab}-\frac{1}{2}Rg_{ab}}\gamma^{ab}
\end{equation*}
and 
\begin{eqnarray}\label{free_grav}
\mathcal{L}_{\textrm{EH}}^{(2)}&\equiv&-\frac{1}{2}\left[\nabla^c\gamma^{ab}\nabla_c\gamma_{ab}
-\nabla^c\gamma\nabla_c\gamma+2\nabla_a\gamma^{ab}\nabla_b\gamma\phantom{\frac{1}{1}}\right.\nonumber\\
&&\phantom{aaaa}-2\nabla^c\gamma^{ab}\nabla_a\gamma_{bc}+R\pp{\gamma^{ab}\gamma_{ab}-\frac{1}{2}\gamma^2}\nonumber\\
&&\left.\phantom{aaaa}-4R^{ab}\pp{\gamma_a\phantom{}^{c}\gamma_{bc}-\frac{1}{2}\gamma\gamma_{ab}}\right],
\end{eqnarray}
where we have defined $\gamma\equiv g^{ab}\gamma_{ab}$. For the scalar field, one obtains
\begin{eqnarray*}
 S_{\Phi}[\phi,g_{ab}+\kappa\gamma_{ab}]&=&
\int_{\mathcal{M}}d^4x\sqrt{-g}[\mathcal{L}_{\Phi}^{(2)}+\mathcal{L}_{\Phi}^{(3)}\\
&&+\mathcal{L}_{\Phi}^{(4)}+\dots],
\end{eqnarray*}
with
\begin{equation}\label{free_field}
\mathcal{L}_{\Phi}^{(2)}\equiv-\frac{1}{2}[\nabla_a\phi\nabla^a\phi+(m^2+\xi R)\phi^2],
\end{equation}
\begin{equation}\label{int_first_order}
\mathcal{L}_{\Phi}^{(3)}\equiv\frac{\kappa}{2}T_{ab}\gamma^{ab},
\end{equation}
and
\begin{eqnarray}\label{int_second_order}
\mathcal{L}_{\Phi}^{(4)}&\equiv&\frac{\kappa^2}{4}[U_{abcd}\gamma^{ab}\gamma^{cd}
+\xi(V_{abcde}\gamma^{ab}\nabla^c\gamma^{de}\nonumber\\
&&\phantom{aa}+W_{abcdef}\nabla^a\gamma^{bc}\nabla^d\gamma^{ef})].
\end{eqnarray}
The tensor $T_{ab}$ appearing in Eq.~(\ref{int_first_order}) was defined in Eq.~(\ref{scalar_field_emt}), while the
expressions for the tensors $U_{abcd}$, $V_{abcde}$, and $W_{abcdef}$ are presented in Eqs.~(\ref{U_tensor}) -
(\ref{W_tensor}) of the Appendix, respectively \cite{footnote_2}. Similarly, the expansion of the classical-matter
action can be written as
\begin{eqnarray*}
 S_{\textrm{M}}[\Psi,g_{ab}+\kappa\gamma_{ab}]&=&
\int_{\mathcal{M}}d^4x\sqrt{-g}[\mathcal{L}_{\textrm{M}}^{(0)}+\mathcal{L}_{\textrm{M}}^{(1)}\\
&&+\mathcal{L}_{\textrm{M}}^{(2)}+\dots],
\end{eqnarray*}
where the specific form of the terms inside the square brackets depends on the Lagrangian assumed for the system. The
background spacetime in which the perturbations are defined will be given by the Einstein equations
\begin{equation}\label{einstein_eq}
 R_{ab}-\frac{1}{2}Rg_{ab}=\frac{\kappa^2}{4} T_{ab}^{\textrm{M}}
\end{equation}
obtained from the zeroth-order action, where $T_{ab}^{\textrm{M}}$ denotes the energy-momentum tensor of the classical
matter. As for the perturbations, it is more convenient to write them in terms of the free scalar field, free graviton,
and interaction actions $S_\phi$, $S_\gamma$, and $S_{\textrm{int}}$, respectively:
\begin{equation}\label{kg_action}
 S_{\phi}[\phi]\equiv\int_{\mathcal{M}}{d^4x\sqrt{-g}\mathcal{L}_{\Phi}^{(2)}},
\end{equation}
\begin{equation}\label{grav_action}
 S_{\gamma}[\gamma_{ab}]\equiv\int_{\mathcal{M}}{d^4x\sqrt{-g}[\mathcal{L}_{\textrm{EH}}^{(2)}
+\mathcal{L}_{\textrm{M}}^{(2)}]},
\end{equation}
and
\begin{equation}\label{int_action}
 S_{\textrm{int}}[\phi,\gamma_{ab}]\equiv\int_{\mathcal{M}}{d^4x\sqrt{-g}[\mathcal{L}_{\Phi}^{(3)}
+\mathcal{L}_{\Phi}^{(4)}]}.
\end{equation}
For the gravitational perturbations, one would need also to specify a gauge to fix the dynamics. However, the
analysis which we will undertake in the next sections dispenses a particular gauge choice. All we have to
assume is that there is a gauge in which the graviton field equation admits stationary oscillatory modes. This
assumption holds, for instance, in the case of a background spacetime curved by a static spherically symmetric star ---
see, e.g., Ref.~\cite{ks_lrr_1999}. 

If the nonminimally coupled free scalar field is destabilized by the curvature of the background spacetime --- like in
the case of the compact object discussed in Sec.\ \ref{sec:gravity_induced_instability} --- the perturbation $\phi$ can
be split into its stable and unstable parts,
\begin{equation}\label{scalar_field_decomposition}
 \phi=\phi_\textrm{s}+\phi_\textrm{u},
\end{equation}
which are defined by 
\begin{equation}\label{stable_sector}
 \phi_\textrm{s}(t,\vt{x})\equiv\sum_{l\mu}\int{d\varpi \phi_\varpi(t) 
F_{\varpi l\mu}(\vt{x})+\textrm{c.c.}}
\end{equation}
and
\begin{equation}\label{unstable_sector}
 \phi_\textrm{u}(t,\vt{x})\equiv \phi_\Omega(t) F_\Omega(\vt{x}).
\end{equation}
By using the orthogonality relations
\begin{equation*}
 \int_\Sigma{d\Sigma F_\Omega(\vt{x})^2}=1,
\end{equation*}
\begin{equation*}
 \int_\Sigma{d\Sigma F_{\varpi l\mu}(\vt{x})F_{\varpi^\prime l^\prime\mu^\prime}(\vt{x})^*}=
\delta_{l l^\prime}\delta_{\mu \mu^\prime}\delta(\varpi-\varpi^\prime),
\end{equation*}
and
\begin{equation*}
 \int_\Sigma{d\Sigma F_\Omega(\vt{x})F_{\varpi l\mu}(\vt{x})^*}=0,
\end{equation*} 
the free scalar field action (\ref{kg_action}) can be cast as
\begin{equation}
 S_\phi[\phi]=S_\phi[\phi_\textrm{s}]+S_\phi[\phi_\textrm{u}].
\end{equation}
For the interaction action, Eq.~(\ref{int_action}), we observe that the tensors $T_{ab}$, $U_{abcd}$, $V_{abcde}$, and
$W_{abcdef}$ depend quadratically on $\phi$ and its derivatives. Hence, by employing the decomposition
(\ref{scalar_field_decomposition}), the energy-momentum tensor can be written as
\begin{equation}\label{scalar_field_emt_decomposition}
 T_{ab}=T_{ab}^\textrm{(s)}+T_{ab}^\textrm{(u)}+t_{ab},
\end{equation}
where $T_{ab}^\textrm{(s/u)}$ corresponds to the tensor given in Eq.~(\ref{scalar_field_emt}) calculated for
$\phi_\textrm{s/u}$ and
\begin{eqnarray}\label{t_tensor}
 t_{ab}&\equiv&(1-2\xi)(\nabla_a\phi_\textrm{s}\nabla_b\phi_\textrm{u}+\nabla_a\phi_\textrm{u}
\nabla_b\phi_\textrm{s})\nonumber\\
&&+2\xi R_{ab}\phi_\textrm{s}\phi_\textrm{u}-2\xi(\phi_\textrm{s}\nabla_a\nabla_b\phi_\textrm{u}\nonumber\\
&&+\phi_\textrm{u}\nabla_a\nabla_b\phi_\textrm{s})+(4\xi-1)[\nabla_c\phi_\textrm{s}\nabla^c\phi_\textrm{u}\nonumber\\
&&+(m^2+\xi R)\phi_\textrm{s}\phi_\textrm{u}]g_{ab}.
\end{eqnarray}
We note that, in contrast to the tensor $T_{ab}^\textrm{(u)}$, the expression given in Eq.~(\ref{t_tensor}) depends
linearly on the amplitude of the field $\phi_\textrm{u}$ and its derivatives. As for the other three tensors, one has
\begin{equation*}
 U_{abcd}=U_{abcd}^\textrm{(s)}+U_{abcd}^\textrm{(u)}+u_{abcd},
\end{equation*}
\begin{equation*}
 V_{abcde}=V_{abcde}^\textrm{(s)}+V_{abcde}^\textrm{(u)}+v_{abcde},
\end{equation*}
and
\begin{equation*}
 W_{abcdef}=W_{abcdef}^\textrm{(s)}+W_{abcdef}^\textrm{(u)}+w_{abcdef}.
\end{equation*}
In these expressions, $U_{abcd}^\textrm{(s/u)}$, $V_{abcde}^\textrm{(s/u)}$, and $W_{abcdef}^\textrm{(s/u)}$ are tensors
obtained from Eqs.~(\ref{U_tensor}) - (\ref{W_tensor}) when one replaces $\phi$ by $\phi_\textrm{s/u}$, while the form
of $u_{abcd}$, $v_{abcde}$, and $w_{abcdef}$ can be easily deduced from these equations and depend linearly on
$\phi_\textrm{s}$ and on $\phi_\textrm{u}$ and its derivatives. We omit the expressions for these tensors, since they
are long and will not contribute in the calculations that follow. This approach of splitting the quantum field into two
sets of modes according to some scale was employed in Ref.~\cite{lm_prd_1996} to study the decoherence of modes above a
certain wavelength in a $\lambda\phi^4$ model, and it is a useful strategy to tackle the issue of the emergence of a
classical order parameter in phase transitions \cite{lmm_prd_2000,lmr_plb_2001,rl_ijtp_2005}.

In our setting, one does not expect quantum fluctuations of the metric and the stable modes of $\phi$ to have any
relevant influence on the background spacetime during the unstable phase. They, however, are perceived by the unstable
mode, becoming entangled with it due to the interaction (\ref{int_action}) in the course of the time evolution.

\subsection{Derivation of the master equation}
\label{subsec:master_eq}

Once one has defined the environment, it is possible to construct the master equation for the reduced density matrix. We
start by assuming that at $t=0$ the total density matrix of the system formed by the scalar perturbations and gravitons
can be written as
\begin{equation*}
 \hat{\varrho}(0)=\hat{\varrho}_\textrm{s}(0)\otimes\hat{\varrho}_\textrm{u}(0)\otimes\hat{\varrho}_\gamma(0),
\end{equation*}
with $\hat{\varrho}_\textrm{s}(0)$, $\hat{\varrho}_\textrm{u}(0)$, and $\hat{\varrho}_\gamma(0)$ denoting the initial
states of the stable modes, unstable mode, and gravitons, respectively. Thus, initially these subsystems are
uncorrelated. In the field amplitude representation, the reduced density matrix for the unstable sector of the field at
$t>0$ is defined by tracing out the gravitons and stable degrees of freedom according to
\begin{equation}\label{reduced_density_matrix}
 \varrho_\textrm{red}(t,\varphi_\textrm{u},\varphi_\textrm{u}\phantom{}^\prime)\equiv
\int{d\varphi_\textrm{s}d\varsigma_{ab}\bra{\varphi_\textrm{u},\varphi_\textrm{s},\varsigma_{ab}}\hat{\varrho}(t)
\ket{\varphi_\textrm{u}\phantom{}^\prime,\varphi_\textrm{s},\varsigma_{ab}}}.
\end{equation}
Above, we have denoted by $\ket{\varphi_\textrm{s}}$, $\ket{\varphi_\textrm{u}}$, and $\ket{\varsigma_{ab}}$ the
eigenstates of the field operators $\hat{\phi}_\textrm{s}$, $\hat{\phi}_\textrm{u}$, and $\hat{\gamma}_{ab}$,
respectively, at $t=0$. The time evolution of the reduced density matrix can be written as
\begin{eqnarray}
 \varrho_\textrm{red}(t,\varphi_\textrm{u},\varphi_\textrm{u}\phantom{}^\prime)&=&
\int d\psi_\textrm{u}d\psi_\textrm{u}\phantom{}^\prime
\varrho_\textrm{u}(0,\psi_\textrm{u},\psi_\textrm{u}\phantom{}^\prime)\nonumber\\
&&\phantom{1}\times J_\textrm{red}(t,\varphi_\textrm{u},\varphi_\textrm{u}\phantom{}^\prime;0,\psi_\textrm{u},
\psi_\textrm{u}\phantom{}^\prime),
\end{eqnarray}
wherein $\varrho_\textrm{u}(0,\psi_\textrm{u},\psi_\textrm{u}\phantom{}^\prime)\equiv
\bra{\psi_\textrm{u}}\hat{\varrho}_\textrm{u}(0)\ket{\psi_\textrm{u}\phantom{}^\prime}$ and $J_\textrm{red}$ stands for
the reduced propagator. Here, the reduced propagator is defined in terms of the following functional integral:
\begin{eqnarray}\label{reduced_propagator}
 J_\textrm{red}(t,\varphi_\textrm{u},\varphi_\textrm{u}\phantom{}^\prime;0,\psi_\textrm{u},
\psi_\textrm{u}\phantom{}^\prime)\equiv
\phantom{1111111111111111111}\nonumber\\
\int_{\psi_\textrm{u}}^{\varphi_\textrm{u}}\mathcal{D}\phi_\textrm{u}
\int_{\psi_\textrm{u}\phantom{}^\prime}^{\varphi_\textrm{u}\phantom{}^\prime}\mathcal{D}\phi_\textrm{u}\phantom{}^\prime
e^{i\{S_\phi[\phi_\textrm{u}]-S_\phi[\phi_\textrm{u}\phantom{}^\prime]\}}F[\phi_\textrm{u},
\phi_\textrm{u}\phantom{}^\prime].
\end{eqnarray}
In Eq.~(\ref{reduced_propagator}), $F$ stands for the Feynman-Vernon influence functional \cite{fv_ap_1963} and is given
by
\begin{eqnarray}\label{feynman_vernon_funct}
 F[\phi_\textrm{u},\phi_\textrm{u}\phantom{}^\prime]&\equiv&
\int d\varphi_\textrm{s}
\int d\psi_\textrm{s}d\psi_\textrm{s}\phantom{}^\prime
\varrho_\textrm{s}(0,\psi_\textrm{s},\psi_\textrm{s}\phantom{}^\prime)\nonumber\\
&&\times\int_{\psi_\textrm{s},\psi_\textrm{s}\phantom{}^\prime}^{\varphi_\textrm{s}=\varphi_\textrm{s}\phantom{}^\prime}
\mathcal{D}\phi_\textrm{s} \mathcal{D}\phi_\textrm{s}\phantom{}^\prime 
e^{i\{S_\phi[\phi_\textrm{s}]-S_\phi[\phi_\textrm{s}\phantom{}^\prime]\}}\nonumber\\
&&\phantom{111111111}\times\tilde{F}[\phi_\textrm{s}+\phi_\textrm{u},
\phi_\textrm{s}\phantom{}^\prime+\phi_\textrm{u}\phantom{}^\prime],
\end{eqnarray}
where we have defined
\begin{eqnarray}
 \tilde{F}[\phi,\phi^\prime]&\equiv&\int d\varsigma_{ab}
\int d\xi_{ab}d\xi_{ab}\phantom{}^\prime
\varrho_\gamma(0,\xi_{ab},\xi_{ab}\phantom{}^\prime)\nonumber\\
&&\times\int_{\xi_{ab},\xi_{ab}\phantom{}^\prime}^{\varsigma_{ab}=\varsigma_{ab}\phantom{}^\prime}
\mathcal{D}\gamma_{ab} \mathcal{D}\gamma_{ab}\phantom{}^\prime 
e^{i\{S_\gamma[\gamma_{ab}]-S_\gamma[\gamma_{ab}\phantom{}^\prime]\}}\nonumber\\
&&\times e^{i\{S_\textrm{int}[\phi,\gamma_{ab}]
-S_\textrm{int}[\phi^\prime,\gamma_{ab}\phantom{}^\prime]\}}.
\end{eqnarray}

The assumption that initially the stable and unstable sectors of the quantum field are uncorrelated is necessary if one
desires to employ the influence functional formalism. This absence of initial correlations would not be the case if the
field, say, had evolved from the vacuum state defined by stationary observers in a previous stable phase. Nevertheless,
initial correlations between the system and its environment are known to affect the dynamics set by the master equation
only in its early stages --- see, e.g.,  Refs.~\cite{rp_pra_1997,zp_leshouches}. This will not be an issue here since in
what follows we will be concerned only with the system in its long-time regime.

Next, we assume that the density matrix $\hat{\varrho}_{\gamma}(0)$ corresponds to a thermal state at temperature $T$
and that $\hat{\varrho}_\textrm{s}(0)=\ket{0_\textrm{s}}\bra{0_\textrm{s}}$ \cite{footnote_3}. By using the closed time
path integral formalism \cite{d_book_1997,ch_book_2008}, we evaluate $\tilde{F}$ up to quadratic order in $\kappa$ and
obtain the following formal expression:
\begin{equation}\label{F_tilde}
 \tilde{F}[\phi,\phi^\prime]=1-\frac{\kappa^2}{4}(I_1[\phi,\phi^\prime]+iI_2[\phi,\phi^\prime]),
\end{equation}
\begin{widetext}
where
\begin{eqnarray}\label{I_1}
 I_1[\phi,\phi^\prime]&\equiv&\int_0^td\tau\int_0^\tau d\tau^\prime 
\int_\Sigma d\Sigma d\Sigma^\prime f(\vt{x})f(\vt{x}^\prime)
\{\textrm{Re}\langle \hat{\gamma}^{ab}(x)\hat{\gamma}^{cd}(x^\prime)\rangle_\beta
[T_{ab}(x)-T^\prime\phantom{}_{ab}(x)]\nonumber\\
&&\times [T_{cd}(x^\prime)-T^\prime\phantom{}_{cd}(x^\prime)]
+i\textrm{Im}\langle \hat{\gamma}^{ab}(x)\hat{\gamma}^{cd}(x^\prime)\rangle_\beta
[T_{ab}(x)-T^\prime\phantom{}_{ab}(x)][T_{cd}(x^\prime)+T^\prime\phantom{}_{cd}(x^\prime)]\},
\end{eqnarray}
and
\begin{eqnarray}\label{I_2}
 I_2[\phi,\phi^\prime]&\equiv&-\int_0^td\tau\int_\Sigma d\Sigma f(\vt{x})
\{\langle\hat{\gamma}^{ab}(x)\hat{\gamma}^{cd}(x)\rangle_\beta
[U_{abcd}(x)-U^\prime\phantom{}_{abcd}(x)]
+\xi \langle\hat{\gamma}^{ab}(x)\nabla^c\hat{\gamma}^{de}(x)\rangle_\beta
[V_{abcde}(x)-V^\prime\phantom{}_{abcde}(x)]\nonumber\\
&&+\xi \langle\nabla^a\hat{\gamma}^{bc}(x)\nabla^d\hat{\gamma}^{ef}(x)\rangle_\beta
[W_{abcdef}(x)-W^\prime\phantom{}_{abcdef}(x)]\}.
\end{eqnarray}
\end{widetext}
In Eqs.~(\ref{I_1}) and (\ref{I_2}), $\langle\dots\rangle_\beta\equiv\textrm{tr}\{\hat{\varrho}_\gamma(0)\dots\}$ is
the thermal average and the tensors $T_{ab}$, $U_{abcd}$, $V_{abcde}$, and $W_{abcdef}$ --- given in
Eqs.~(\ref{scalar_field_emt}) and (\ref{U_tensor})-(\ref{W_tensor}) --- are calculated for $\phi$, while
$T^\prime\phantom{}_{ab}$, $U^\prime\phantom{}_{abcd}$, $V^\prime\phantom{}_{abcde}$, and $W^\prime\phantom{}_{abcdef}$
are calculated for $\phi^\prime$. The functional $I_2$ is clearly divergent and must be absorbed into the bare
parameters of the free scalar field action, Eq.~(\ref{kg_action}). We shall not delve into the question of what
corrections this procedure may introduce here, since it does not contribute to decoherence effects --- for a discussion
on the quantum corrections induced by a thermal bath of gravitons in flat spacetime, see, e.g.,
Ref.~\cite{apv_prd_2004}. Thus, we are left with
\begin{equation}\label{F_tilde_renormalized}
 \tilde{F}[\phi,\phi^\prime]=1-\frac{\kappa^2}{4}I_1[\phi,\phi^\prime].
\end{equation}
Then, by substituting Eq.~(\ref{F_tilde_renormalized}) into Eq.~(\ref{feynman_vernon_funct}) and using
Eq.~(\ref{scalar_field_emt_decomposition}), one obtains
\begin{widetext}
\begin{equation}\label{renormalized_feynman_vernon_funct}
 F[\phi_\textrm{u},\phi_\textrm{u}\phantom{}^\prime]=1-\frac{\kappa^2}{4}
(G_1[\phi_\textrm{u},\phi_\textrm{u}\phantom{}^\prime]
+G_2[\phi_\textrm{u},\phi_\textrm{u}\phantom{}^\prime]
+G_3[\phi_\textrm{u},\phi_\textrm{u}\phantom{}^\prime]),
\end{equation}
with
\begin{equation}\label{G_1}
 G_1[\phi_\textrm{u},\phi_\textrm{u}\phantom{}^\prime]\equiv
2i\int_0^t d\tau\int_0^\tau d\tau^\prime\int_\Sigma d\Sigma d\Sigma^\prime
f(\vt{x})f(\vt{x}^\prime)
\textrm{Im}\langle \hat{\gamma}^{ab}(x)\hat{\gamma}^{cd}(x^\prime)\rangle_\beta
\langle \hat{T}^\textrm{(s)}_{cd}(x^\prime)\rangle_0
[T^{(\textrm{u})}_{ab}(x)-T^{(\textrm{u})\prime}_{ab}(x)],
\end{equation}
\begin{eqnarray}\label{G_2}
 G_2[\phi_\textrm{u},\phi_\textrm{u}\phantom{}^\prime]&\equiv&
\int_0^td\tau\int_0^\tau d\tau^\prime 
\int_\Sigma d\Sigma d\Sigma^\prime f(\vt{x})f(\vt{x}^\prime)
\{\textrm{Re}\langle \hat{\gamma}^{ab}(x)\hat{\gamma}^{cd}(x^\prime)\rangle_\beta
[T^{(\textrm{u})}_{ab}(x)-T^{(\textrm{u})\prime}_{ab}(x)]
[T^{(\textrm{u})}_{cd}(x^\prime)-T^{(\textrm{u})\prime}_{cd}(x^\prime)]\nonumber\\
&&+i\textrm{Im}\langle \hat{\gamma}^{ab}(x)\hat{\gamma}^{cd}(x^\prime)\rangle_\beta
[T^{(\textrm{u})}_{ab}(x)-T^{(\textrm{u})\prime}_{ab}(x)]
[T^{(\textrm{u})}_{cd}(x^\prime)+T^{(\textrm{u})\prime}_{cd}(x^\prime)]\},
\end{eqnarray}
and
\begin{eqnarray}\label{G_3}
 G_3[\phi_\textrm{u},\phi_\textrm{u}\phantom{}^\prime]&\equiv&
\int d\varphi_\textrm{s}\int d\psi_\textrm{s}
d\psi_\textrm{s}\phantom{}^\prime\varrho_\textrm{s}(0,\psi_\textrm{s},\psi_\textrm{s}\phantom{}^\prime)
\int_{\psi_\textrm{s},\psi_\textrm{s}\phantom{}^\prime}^{\varphi_\textrm{s}=\varphi_\textrm{s}\phantom{}^\prime}
\mathcal{D}\phi_\textrm{s} \mathcal{D}\phi_\textrm{s}\phantom{}^\prime 
e^{i\{S_\phi[\phi_\textrm{s}]-S_\phi[\phi_\textrm{s}\phantom{}^\prime]\}}
\int_0^td\tau\int_0^\tau d\tau^\prime 
\int_\Sigma d\Sigma d\Sigma^\prime f(\vt{x})f(\vt{x}^\prime)\nonumber\\
&&\times\{\textrm{Re}\langle \hat{\gamma}^{ab}(x)\hat{\gamma}^{cd}(x^\prime)\rangle_\beta
[t_{ab}(x)-t^\prime\phantom{}_{ab}(x)]
[t_{cd}(x^\prime)-t^\prime\phantom{}_{cd}(x^\prime)]
+i\textrm{Im}\langle \hat{\gamma}^{ab}(x)\hat{\gamma}^{cd}(x^\prime)\rangle_\beta
[t_{ab}(x)-t^\prime\phantom{}_{ab}(x)]\nonumber\\
&&\times [t_{cd}(x^\prime)+t^\prime\phantom{}_{cd}(x^\prime)]\}.
\end{eqnarray}
\end{widetext}
In Eq.~(\ref{G_1}), $\langle \hat{T}^\textrm{(s)}_{ab}\rangle_0$ stands for the renormalized expectation value in the
state $\ket{0_\textrm{s}}$ of the energy-momentum tensor operator associated with $\hat{\phi}_\textrm{s}$. As for the
tensors $T^{(\textrm{u})}_{ab}$ and $t_{ab}$ in Eqs.~(\ref{G_1}) - (\ref{G_3}), they are constructed from
$\phi_\textrm{u}$ and $\phi_\textrm{s}$, while $T^{(\textrm{u})\prime}_{ab}$ and $t^\prime\phantom{}_{ab}$ are
constructed from $\phi_\textrm{u}\phantom{}^\prime$ and $\phi_\textrm{s}\phantom{}^\prime$. 

The terms in the influence functional (\ref{renormalized_feynman_vernon_funct}) responsible for decoherence effects and
damping are those given in Eqs.~(\ref{G_2}) and (\ref{G_3}). (Of course, these functionals depend on the specific
interaction of the unstable mode with its environment.) While the former comes from the interaction of the unstable mode
with gravitons through its energy-momentum tensor, the latter is a consequence of the interaction of the unstable mode
with the whole environment via its amplitude and derivatives. Consequently, one expects that under the influence of the
functional (\ref{G_2}), the density matrix will tend to evolve into a mixture of states which bear some relation with
the energy-momentum tensor operator. The functional (\ref{G_3}), on the other hand, will tend to diagonalize the density
matrix in the basis of localized states in amplitude and momentum of the unstable mode. The set of states in which the
density matrix becomes diagonal is known in the literature as ``pointer states.'' Pointer states are those states
less affected by the environment; i.e., they are the states less willing to evolve into an entangled state with the
environment \cite{gjkksz_book_1996,z_prd_1981_1982}. The implications of the terms in Eq.~(\ref{G_2}) were recently
investigated in Ref.~\cite{b_prl_2013} for the case of a flat spacetime background. There it was shown that in the
nonrelativistic regime the density matrix tends to become diagonal in the energy basis --- see also
Ref.~\cite{ah_cqg_2013}. Here, however, we will be concerned with the decoherence effects introduced by terms in
Eq.~(\ref{G_3}) in the full relativistic curved spacetime regime.

In order to obtain the master equation for the density matrix (\ref{reduced_density_matrix}), we need to calculate the
time derivative of the reduced propagator $J_\textrm{red}$. The form of the propagator can be computed by applying the
saddle point approximation to the functional integral in Eq.~(\ref{reduced_propagator}). In this approximation, one has
for $J_\textrm{red}$ that 
\begin{equation}\label{saddle_point_reduced_propagator}
 J_\textrm{red}(t,\varphi_\textrm{u},\varphi_\textrm{u}\phantom{}^\prime;
0,\psi_\textrm{u},\psi_\textrm{u}\phantom{}^\prime)
\approx\exp\{iA[\phi_\textrm{u}^\textrm{cl},\phi_\textrm{u}^{\textrm{cl}\prime}]\},
\end{equation}
with the total effective action
\begin{equation*}
 A[\phi_\textrm{u},\phi_\textrm{u}\phantom{}^\prime]\equiv
S_\phi[\phi_\textrm{u}]-S_\phi[\phi_\textrm{u}\phantom{}^\prime]
+S_\textrm{IF}[\phi_\textrm{u},\phi_\textrm{u}\phantom{}^\prime],
\end{equation*}
and the influence action $S_\textrm{IF}$ being implicitly defined through
$F[\phi_\textrm{u},\phi_\textrm{u}\phantom{}^\prime]=\exp\{iS_\textrm{IF}[\phi_\textrm{u},\phi_\textrm{u}\phantom{}
^\prime]\}$. Above, $\phi_\textrm{u}^\textrm{cl}$ and $\phi_\textrm{u}^{\textrm{cl}\prime}$ are solutions of the
equation of motion,
\begin{equation*}
\left.\frac{\delta\textrm{Re}A}{\delta\phi_\textrm{u}}\right|_{\phi_\textrm{u}\phantom{}^\prime=\phi_\textrm{u}}=0,
\end{equation*}
satisfying the conditions $\phi_\textrm{u}^\textrm{cl}(0,\vt{x})=\psi_\textrm{u}(\vt{x})$,
$\phi_\textrm{u}^\textrm{cl}(t,\vt{x})=\varphi_\textrm{u}(\vt{x})$,
$\phi_\textrm{u}^{\textrm{cl}\prime}(0,\vt{x})=\psi_\textrm{u}\phantom{}^\prime(\vt{x})$, and
$\phi_\textrm{u}^{\textrm{cl}\prime}(t,\vt{x})=\varphi_\textrm{u}\phantom{}^\prime(\vt{x})$. 

At this point, it is clear that decoherence will be induced by terms in the propagator $J_\textrm{red}$ which are of
order $\kappa^2$. Consequently, one can approximate $\phi_\textrm{u}^\textrm{cl}$ and
$\phi_\textrm{u}^{\textrm{cl}\prime}$ by unstable solutions of the free scalar field equation with the appropriate
conditions at the initial and final instants. Thus, the classical solutions will have the form  
\begin{equation}\label{class_sol}
 \phi_\textrm{u}^\textrm{cl}(\tau,\vt{x})=\phi_\Omega^\textrm{cl}(\tau)F_\Omega(\vt{x}),
\end{equation}
with
\begin{equation}\label{class_sol_temporal_part}
\phi_\Omega^\textrm{cl}(\tau)\equiv q_0\frac{\sinh\Omega(t-\tau)}{\sinh\Omega t}
+q\frac{\sinh\Omega\tau}{\sinh\Omega t}.
\end{equation}
Now, inserting Eq.~(\ref{class_sol}) into the action (\ref{kg_action}), one obtains
\begin{eqnarray}\label{classical_action}
 S_\phi[\phi_\textrm{u}^\textrm{cl}]&=&\frac{1}{2}\int_0^t{d\tau[(\dot{\phi}_\Omega^{\textrm{cl}})^2
+\Omega^2(\phi_\Omega^{\textrm{cl}})^2]}\nonumber\\
&=&\frac{\Omega}{2\sinh\Omega t}[(q_0^2+q^2)\cosh\Omega t-2q_0q],\\
\nonumber
\end{eqnarray}
with the second equality above coming from Eq.~(\ref{class_sol_temporal_part}). As for the tensors appearing in
Eqs.~(\ref{G_1}) - (\ref{G_3}), we shall denote by $T_{ab}^\textrm{cl}$ and $t_{ab}^\textrm{cl}$ the field
energy-momentum tensor and the tensor given in Eq.~(\ref{t_tensor}), respectively, when calculated for
$\phi_\textrm{u}^\textrm{cl}$. By employing Eq.~(\ref{class_sol}), the expression for $t_{ab}^\textrm{cl}$ can be cast
as
\begin{widetext}
\begin{equation}\label{t_cl}
 t_{ab}^\textrm{cl}=r_{ab}^{(1)}\dot{\phi}_\Omega^\textrm{cl}+r_{ab}^{(2)}\phi_\Omega^\textrm{cl},
\end{equation}
where
\begin{equation}\label{r_1}
 r_{ab}^{(1)}\equiv\frac{F_\Omega}{\sqrt{f}}\{[n_an_b+(1-4\xi)h_{ab}]n^c\nabla_c\phi_\textrm{s}
+2(2\xi-1) n_{(a}D_{b)}\phi_\textrm{s}-4\xi n_{(a}a_{b)}\phi_\textrm{s}\}
+4\xi\frac{\phi_\textrm{s}}{\sqrt{f}}n_{(a}D_{b)}F_\Omega,
\end{equation}
\begin{eqnarray}\label{r_2}
 r_{ab}^{(2)}&\equiv&n_an_b\{(1-4\xi)D^cF_\Omega D_c\phi_\textrm{s}+(m^2+\xi R)F_\Omega\phi_\textrm{s}
+2\xi(D_ca^c+a_ca^c)F_\Omega\phi_\textrm{s}-2\xi F_\Omega D_cD^c\phi_\textrm{s}\nonumber\\
&&-2\xi D_cD^cF_\Omega\phi_\textrm{s}\}-2(1-2\xi)n_{(a}D_{b)}F_\Omega n^c\nabla_c\phi_\textrm{s}
+4\xi F_\Omega n_{(a}D_{b)}(n^c\nabla_c\phi_\textrm{s})+2(1-2\xi)D_{(a}F_\Omega D_{b)}\phi_\textrm{s}\nonumber\\
&&-(1-4\xi)h_{ab}[D_cF_\Omega D^c\phi_\textrm{s}+(m^2+\xi R)F_\Omega \phi_\textrm{s}]
-2\xi(F_\Omega D_aD_b\phi_\textrm{s}+D_aD_bF_\Omega\phi_\textrm{s})\nonumber\\
&&+2\xi[\phantom{}^{(3)}R_{ab}-(D_aa_b+a_aa_b)]F_\Omega\phi_\textrm{s},
\end{eqnarray}
and we recall that $f$ was defined in Eq.~(\ref{star_metric}) and $n^a=f^{-1/2}(\del_t)^a$. In Eqs.~(\ref{r_1})
and (\ref{r_2}), $a^a\equiv n^c\nabla_cn^a$ is the acceleration of the observers following the orbits of the timelike
Killing vector field, $D_a$ is the derivative operator associated with the spatial metric $h_{ab}$, and
$\phantom{}^{(3)}R_{ab}$ denotes the Ricci tensor of the spatial section. Finally, by combining
Eqs.~(\ref{renormalized_feynman_vernon_funct}) and (\ref{saddle_point_reduced_propagator}) and then using
Eqs.~(\ref{classical_action}) and (\ref{t_cl}), one obtains an expression for $J_\textrm{red}$.

For the calculation of the reduced propagator time derivative in the saddle point approximation, it is useful to
define the operators $\hat{P}$ and $\hat{Q}$ as
\begin{equation}\label{P_operator}
 \hat{P}(\tau)\equiv\int_\Sigma{d\Sigma f(\vt{x})\hat{\gamma}^{ab}(\tau,\vt{x})\hat{r}_{ab}^{(1)}(\tau,\vt{x})}
\end{equation}
and
\begin{equation}\label{Q_operator}
 \hat{Q}(\tau)\equiv\int_\Sigma{d\Sigma f(\vt{x})\hat{\gamma}^{ab}(\tau,\vt{x})\hat{r}_{ab}^{(2)}(\tau,\vt{x})},
\end{equation}
where $\hat{r}_{ab}^{(1)}$ and $\hat{r}_{ab}^{(2)}$ are obtained from Eqs.~(\ref{r_1}) and (\ref{r_2}) by replacing
$\phi_\textrm{s}$ by the field operator $\hat{\phi}_\textrm{s}$. Then, the time derivative of the reduced propagator
can be written as
\begin{eqnarray}\label{reduced_propagator_time_derivative}
 \del_t J_\textrm{red}&=&
\del_tJ_0+\Bigg{\{}2i\int_0^t d\tau^\prime \int_\Sigma d\Sigma d\Sigma^\prime f(\vt{x})f(\vt{x}^\prime)
D^{abcd}(t,\vt{x};\tau^\prime,\vt{x}^\prime)\langle\hat{T}^\textrm{(s)}_{cd}(\tau^\prime,\vt{x}^\prime)\rangle_0
[T^{\textrm{cl}}_{ab}(t,\vt{x})-T^{\textrm{cl}\prime}_{ab}(t,\vt{x})]\nonumber\\
&&-\int_0^td\tau^\prime \int_\Sigma d\Sigma d\Sigma^\prime f(\vt{x})f(\vt{x}^\prime)
\{N^{abcd}(t,\vt{x};\tau^\prime,\vt{x}^\prime)
[T^{\textrm{cl}}_{ab}(t,\vt{x})-T^{\textrm{cl}\prime}_{ab}(t,\vt{x})]
[T^{\textrm{cl}}_{cd}(\tau^\prime,\vt{x}^\prime)-T^{\textrm{cl}\prime}_{cd}(\tau^\prime,\vt{x}^\prime)]\nonumber\\
&&-iD^{abcd}(t,\vt{x};\tau^\prime,\vt{x}^\prime)
[T^{\textrm{cl}}_{ab}(t,\vt{x})-T^{\textrm{cl}\prime}_{ab}(t,\vt{x})]
[T^{\textrm{cl}}_{cd}(\tau^\prime,\vt{x}^\prime)+T^{\textrm{cl}\prime}_{cd}(\tau^\prime,\vt{x}^\prime)]\}
-\frac{\kappa^2}{4}\int_0^td\tau^\prime\{\textrm{Re}\langle\hat{Q}(t)\hat{Q}(\tau^\prime)\rangle\nonumber\\
&&\times[\phi_\Omega^\textrm{cl}(t)-\phi_\Omega^{\textrm{cl}\prime}(t)]
[\phi_\Omega^\textrm{cl}(\tau^\prime)-\phi_\Omega^{\textrm{cl}\prime}(\tau^\prime)]
+\textrm{Re}\langle\hat{P}(t)\hat{P}(\tau^\prime)\rangle 
[\dot{\phi}_\Omega^\textrm{cl}(t)-\dot{\phi}_\Omega^{\textrm{cl}\prime}(t)]
[\dot{\phi}_\Omega^\textrm{cl}(\tau^\prime)-\dot{\phi}_\Omega^{\textrm{cl}\prime}(\tau^\prime)]\nonumber\\
&&+\textrm{Re}\langle\hat{Q}(t)\hat{P}(\tau^\prime)\rangle 
[\phi_\Omega^\textrm{cl}(t)-\phi_\Omega^{\textrm{cl}\prime}(t)]
[\dot{\phi}_\Omega^\textrm{cl}(\tau^\prime)-\dot{\phi}_\Omega^{\textrm{cl}\prime}(\tau^\prime)]
+\textrm{Re}\langle\hat{P}(t)\hat{Q}(\tau^\prime)\rangle
[\dot{\phi}_\Omega^\textrm{cl}(t)-\dot{\phi}_\Omega^{\textrm{cl}\prime}(t)]
[\phi_\Omega^\textrm{cl}(\tau^\prime)-\phi_\Omega^{\textrm{cl}\prime}(\tau^\prime)]\nonumber\\
&&+i\textrm{Im}\langle\hat{Q}(t)\hat{Q}(\tau^\prime)\rangle 
[\phi_\Omega^\textrm{cl}(t)-\phi_\Omega^{\textrm{cl}\prime}(t)]
[\phi_\Omega^\textrm{cl}(\tau^\prime)+\phi_\Omega^{\textrm{cl}\prime}(\tau^\prime)]
+i\textrm{Im}\langle\hat{P}(t)\hat{P}(\tau^\prime)\rangle 
[\dot{\phi}_\Omega^\textrm{cl}(t)-\dot{\phi}_\Omega^{\textrm{cl}\prime}(t)]
[\dot{\phi}_\Omega^\textrm{cl}(\tau^\prime)+\dot{\phi}_\Omega^{\textrm{cl}\prime}(\tau^\prime)]\nonumber\\
&&+i\textrm{Im}\langle\hat{Q}(t)\hat{P}(\tau^\prime)\rangle 
[\phi_\Omega^\textrm{cl}(t)-\phi_\Omega^{\textrm{cl}\prime}(t)]
[\dot{\phi}_\Omega^\textrm{cl}(\tau^\prime)+\dot{\phi}_\Omega^{\textrm{cl}\prime}(\tau^\prime)]
+i\textrm{Im}\langle\hat{P}(t)\hat{Q}(\tau^\prime)\rangle
[\dot{\phi}_\Omega^\textrm{cl}(t)-\dot{\phi}_\Omega	^{\textrm{cl}\prime}(t)]
[\phi_\Omega^\textrm{cl}(\tau^\prime)+\phi_\Omega^{\textrm{cl}\prime}(\tau^\prime)]\}\Bigg{\}}\nonumber\\
&&\times J_0,
\end{eqnarray}
with
\begin{equation*}
 N^{abcd}(\tau,\vt{x};\tau^\prime,\vt{x})\equiv \frac{\kappa^2}{4}
\textrm{Re}\langle\hat{\gamma}^{ab}(\tau,\vt{x})\hat{\gamma}^{cd}(\tau^\prime,\vt{x}^\prime)\rangle_\beta
\end{equation*}
and
\begin{equation*}
 D^{abcd}(\tau,\vt{x};\tau^\prime,\vt{x})\equiv -\frac{\kappa^2}{4}
\textrm{Im}\langle\hat{\gamma}^{ab}(\tau,\vt{x})\hat{\gamma}^{cd}(\tau^\prime,\vt{x}^\prime)\rangle_\beta,
\end{equation*}
while $\langle\dots\rangle\equiv\textrm{tr}\{\hat{\varrho}_\textrm{s}(0)\otimes\hat{\varrho}_\gamma(0)\dots\}$ and
$J_0$ denotes the free propagator for the unstable mode. The free propagator is defined by
\begin{eqnarray}\label{free_propagator}
 J_0(t,\varphi_\textrm{u},\varphi_\textrm{u}\phantom{}^\prime;0,\psi_\textrm{u},
\psi_\textrm{u}\phantom{}^\prime)\equiv
\int_{\psi_\textrm{u}}^{\varphi_\textrm{u}}\mathcal{D}\phi_\textrm{u}
\int_{\psi_\textrm{u}\phantom{}^\prime}^{\varphi_\textrm{u}\phantom{}^\prime}\mathcal{D}\phi_\textrm{u}\phantom{}
^\prime e^{i\{S_\phi[\phi_\textrm{u}]-S_\phi[\phi_\textrm{u}\phantom{}^\prime]\}},
\end{eqnarray}
and in terms of the initial and final amplitudes of the unstable mode, it can be cast as
\begin{equation*}
 J_0\propto \exp\left\{\frac{i\Omega}{2\sinh\Omega t}[(q^2-q^{\prime2}+q_0^2-q_0^{\prime2})\cosh\Omega t 
-2(q_0q-q_0^\prime q^\prime)]\right\}.
\end{equation*}
The last expression implies that $J_0$ satisfies the following relations:
\begin{equation}\label{identity_1}
 \phi_\Omega^\textrm{cl}(\tau)J_0=\cc{\cosh\Omega(t-\tau)q+\frac{\sinh\Omega(t-\tau)}{\Omega}i\del_q}J_0
\end{equation}
and 
\begin{equation}\label{identity_2}
 \phi_\Omega^{\textrm{cl}\prime}(\tau)J_0=\cc{\cosh\Omega(t-\tau)q^\prime
-\frac{\sinh\Omega(t-\tau)}{\Omega}i\del_{q^\prime}}J_0.
\end{equation}
Consequently, one can employ Eqs.~(\ref{identity_1}) and (\ref{identity_2}) to obtain the master equation for
$\varrho_\textrm{red}$:
\begin{eqnarray}\label{master_eq}
 \del_t\varrho_\textrm{red}&=&-i\cc{\frac{1}{2}(-\del_q^2+\del_{q^\prime}^2)-\frac{\Omega^2}{2}(q^2-q^{\prime2})}
\varrho_\textrm{red}\nonumber\\
&&-\frac{\kappa^2}{4}\int_0^td\tau[\textrm{Re}\langle\hat{Q}(\tau)\hat{Q}(0)\rangle\cosh\Omega\tau
-\Omega\textrm{Re}\langle\hat{Q}(\tau)\hat{P}(0)\rangle\sinh\Omega\tau](q-q^\prime)^2\varrho_\textrm{red}\nonumber\\
&&-\frac{\kappa^2}{4}\int_0^td\tau[\textrm{Re}\langle\hat{P}(\tau)\hat{P}(0)\rangle\cosh\Omega\tau
-\Omega^{-1}\textrm{Re}\langle\hat{P}(\tau)\hat{Q}(0)\rangle\sinh\Omega\tau](-i\del_q-i\del_{q^\prime})^2
\varrho_\textrm{red}\nonumber\\
&&+\frac{\kappa^2}{4}\int_0^td\tau\{[\Omega^{-1}\textrm{Re}\langle\hat{Q}(\tau)\hat{Q}(0)\rangle
+\Omega\textrm{Re}\langle\hat{P}(\tau)\hat{P}(0)\rangle]\sinh\Omega\tau\nonumber\\
&&-[\textrm{Re}\langle\hat{P}(\tau)\hat{Q}(0)\rangle
+\textrm{Re}\langle\hat{Q}(\tau)\hat{P}(0)\rangle]\cosh\Omega\tau\}(q-q^\prime)(-i\del_q-i\del_{q^\prime})
\varrho_\textrm{red}+\dots.
\end{eqnarray}
The temporal arguments of the factors inside the integrals above were rearranged using the fact that both
$\hat{\varrho}_\textrm{s}(0)$ and $\hat{\varrho}_\gamma(0)$ are stationary states. As for the Wigner function
$W_\textrm{red}$ associated with the state $\hat{\varrho}_\textrm{red}$, one obtains from the master equation above that
it satisfies
\begin{eqnarray}\label{model_wf_eq}
 \del_t W_\textrm{red}&=&\{H(q,p),W_\textrm{red}\}
+\frac{\kappa^2}{4}\int_0^td\tau[\textrm{Re}\langle\hat{Q}(\tau)\hat{Q}(0)\rangle\cosh\Omega\tau
-\Omega\textrm{Re}\langle\hat{Q}(\tau)\hat{P}(0)\rangle\sinh\Omega\tau]\del_p^2W_\textrm{red}\nonumber\\
&&+\frac{\kappa^2}{4}\int_0^td\tau[\textrm{Re}\langle\hat{P}(\tau)\hat{P}(0)\rangle\cosh\Omega\tau
-\Omega^{-1}\textrm{Re}\langle\hat{P}(\tau)\hat{Q}(0)\rangle\sinh\Omega\tau]\del_q^2
W_\textrm{red}\nonumber\\
&&+\frac{\kappa^2}{4}\int_0^td\tau\{[\Omega^{-1}\textrm{Re}\langle\hat{Q}(\tau)\hat{Q}(0)\rangle
+\Omega\textrm{Re}\langle\hat{P}(\tau)\hat{P}(0)\rangle]\sinh\Omega\tau\nonumber\\
&&-[\textrm{Re}\langle\hat{P}(\tau)\hat{Q}(0)\rangle
+\textrm{Re}\langle\hat{Q}(\tau)\hat{P}(0)\rangle]\cosh\Omega\tau\}\del_p\del_q
W_\textrm{red}+\dots.
\end{eqnarray}
\end{widetext}
In the master equation (\ref{master_eq}), we present only its free dynamics term and the members engendered by the
interaction between the stable modes and the unstable one which are able to cause loss of quantum coherence, while the
ellipsis encloses all the other terms. The terms written explicitly in the right-hand side of Eq.~(\ref{master_eq})
tend to localize the state of the unstable mode both in amplitude and momentum representations. This resembles the
problem of localization of particles \cite{jz_zpb_1985} and the analysis of the quantum Brownian motion problem
\cite{cl_physa_1983,uz_prd_1989,hpz_prd_1992_1993}. As for the terms originated by the direct interaction between the
unstable mode and gravitons, they depend quadratically on the amplitude and momentum and, thus, are not expected to
localize the state in these representations.

We note that, even though it was assumed that the background is curved by a compact spherical object, the analysis we
carried out so far applies for more general static spacetimes.

\subsection{Long-time regime}
\label{subsec:long_time_regime}

Our next task is to show that Eq.~(\ref{master_eq}) does localize the state of the unstable mode in the amplitude and
momentum representations. To do so, let us assume that initially the unstable mode is in the state $\ket{0_\textrm{u}}$.
We emphasize that this choice only serves to simplify the calculations. Due to the feebleness of the gravitational
interaction, the evolution dictated by the master equation (\ref{master_eq}) is dominated by the free field evolution.
As discussed in Sec.\ \ref{sec:free_evolution}, the free dynamics acts as a squeeze operator. Thus, in the long-time
regime ($\Omega t\gg1$) the initial state becomes highly squeezed. In this case Eq.~(\ref{squeezed_vacuum_eq}) leads to
\begin{equation}\label{squeezed_vacuum_eq_amplitude_rep_1}
 -i\del_q\varrho_\textrm{red}\approx\Omega q\varrho_\textrm{red}
\end{equation}
and 
\begin{equation}\label{squeezed_vacuum_eq_amplitude_rep_2}
 i\del_{q^\prime}\varrho_\textrm{red}\approx\Omega q^\prime\varrho_\textrm{red}.
\end{equation}
Then, by substituting Eqs.~(\ref{squeezed_vacuum_eq_amplitude_rep_1}) and (\ref{squeezed_vacuum_eq_amplitude_rep_2})
into Eq.~(\ref{master_eq}), the master equation reduces to
\begin{eqnarray}\label{long_time_regime_master_eq}
 \del_t\varrho_\textrm{red}&\approx&-i\cc{\frac{1}{2}(-\del_q^2+\del_{q^\prime}^2)-\frac{\Omega^2}{2}(q^2-q^{\prime2})}
\varrho_\textrm{red}\nonumber\\
&&-D(q-q^\prime)^2\varrho_\textrm{red}+\dots,
\end{eqnarray}
with the diffusion coefficient $D>0$ given by
\begin{eqnarray}\label{diff_coefficient}
 D&\equiv&\frac{\kappa^2}{4}\int_0^{+\infty}d\tau K(\tau)e^{-\Omega\tau}
\end{eqnarray}
and
\begin{equation}\label{K}
 K(\tau-\tau^\prime)\equiv\textrm{Re}\langle[\hat{Q}(\tau)+\Omega\hat{P}(\tau)]
[\hat{Q}(\tau^\prime)+\Omega\hat{P}(\tau^\prime)]\rangle.
\end{equation}
As for $W_\textrm{red}$, one has the following equation:
\begin{equation}\label{long_time_regime_wf_eq}
 \del_t W_\textrm{red}\approx\{H(q,p),W_\textrm{red}\}+D\del_p^2 W_\textrm{red}+\dots.
\end{equation}
The second term in Eq.~(\ref{long_time_regime_master_eq}) ensures the localization of $\hat{\varrho}_\textrm{red}$ in
both amplitude and momentum representations, due to the relation between $q$ and $p$ set by
Eqs.~(\ref{squeezed_vacuum_eq_amplitude_rep_1}) and (\ref{squeezed_vacuum_eq_amplitude_rep_2}).

One can estimate the magnitude of the diffusion coefficient in the following manner. As mentioned earlier, we assume
the existence of a gauge in which the graviton field admits the following decomposition:
\begin{equation}\label{grav_operator}
 \hat{\gamma}_{ab}(t,\vt{x})=\sum_j\int\frac{d\vartheta(\alpha)}{\sqrt{2\omega_\alpha}}\hat{b}_\alpha^{(j)}
e^{-i\omega_\alpha t}\varepsilon_{\alpha ab}^{(j)}(\vt{x})+\textrm{H.c.}
\end{equation}
Here, $\alpha$ denotes all the pertinent quantum numbers, $j$ labels the graviton polarizations,
$\hat{b}_\alpha^{(j)}$ denotes the graviton annihilation operator, $\varepsilon_{\alpha ab}^{(j)}$ is the
spatial part of the mode, and $\omega_\alpha>0$. As for the field operator $\hat{\phi}_\textrm{s}$, one has
\begin{equation}\label{stable_sector_field_operator}
 \hat{\phi}_\textrm{s}(t,\vt{x})=\sum_{l\mu}\int d\varpi\hat{a}_{\varpi l\mu}
v_{\varpi l\mu}^{(+)}(t,\vt{x})+\textrm{H.c.},
\end{equation}
with $v_{\varpi l\mu}^{(+)}$ given in Eq.~(\ref{v_modes}). Thus, by substituting Eqs.~(\ref{grav_operator}) and
(\ref{stable_sector_field_operator}) into the expressions for the operators $\hat{P}$ and $\hat{Q}$,
Eqs.~(\ref{P_operator}) and (\ref{Q_operator}), with the aid of Eqs.~(\ref{r_1}) and (\ref{r_2}), Eq.~(\ref{K}) can be
cast as
\begin{eqnarray}\label{K_expression}
 K(\tau-\tau^\prime)&=&\textrm{Re}
\sum_{jl\mu}\int \frac{d\vartheta(\alpha)}{2\omega_\alpha}
\frac{d\varpi}{2\varpi}\left[e^{-i(\omega_\alpha+\varpi)(\tau-\tau^\prime)}\phantom{\frac{e^{i}}{e^{i}}|^2}
\right.\nonumber\\
&&\times\frac{e^{\beta\omega_\alpha}}{e^{\beta\omega_\alpha}-1}
|\phantom{}_1\Gamma_{l\mu}^{(j)}(\alpha,\varpi)|^2\nonumber\\
&&\left.+\frac{e^{i(\omega_\alpha-\varpi)(\tau-\tau^\prime)}}{e^{\beta\omega_\alpha}-1}
|\phantom{}_2\Gamma_{l\mu}^{(j)}(\alpha,\varpi)|^2\right].
\end{eqnarray}
In Eq.~(\ref{K_expression}), $\beta\equiv(k_\textrm{B}T)^{-1}$,
\begin{equation}\label{Gamma_1}
 \phantom{}_1\Gamma_{l\mu}^{(j)}(\alpha,\varpi)\equiv\int_\Sigma d\Sigma f
\varepsilon_{\alpha ab}^{(j)}[\Omega s_{\varpi l\mu}^{(1)ab}+s_{\varpi l\mu}^{(2)ab}]
\end{equation}
and
\begin{equation}\label{Gamma_2}
 \phantom{}_2\Gamma_{l\mu}^{(j)}(\alpha,\varpi)\equiv\int_\Sigma d\Sigma f
\varepsilon_{\alpha ab}^{(j)*}[\Omega s_{\varpi l\mu}^{(1)ab}+s_{\varpi l\mu}^{(2)ab}],
\end{equation}
where $s_{\varpi l\mu}^{(1)ab}$ and $s_{\varpi l\mu}^{(2)ab}$ come from Eqs.~(\ref{r_1}) and (\ref{r_2}),
respectively, after we replace $\phi_\textrm{s}$ by $v_{\varpi l\mu}^{(+)}$ and factorize $e^{-i\varpi\tau}$. The
tensors $s_{\varpi l\mu}^{(1)ab}$ and $s_{\varpi l\mu}^{(2)ab}$ are weighted by the spatial part of the unstable mode,
$F_\Omega=\psi_\Omega Y_{00}/r$, with $\psi_\Omega$ as a ``bound solution'' of Eq.~(\ref{psi_Omega_eq}) with width of
order $\Omega^{-1}$. As a result, $\phantom{}_1\Gamma_{l\mu}^{(j)}$ and $\phantom{}_2\Gamma_{l\mu}^{(j)}$ can be
neglected for $\alpha$ and $\varpi$ such that $\omega_\alpha,\varpi\gg\Omega$. Thus, the main contribution for the
integrals in Eq.~(\ref{K_expression}) comes from gravitons and stable modes with frequency up to order $\Omega$.
Consequently, one can define the high-temperature regime here as $k_\textrm{B}T\gg\Omega$. In the spacetime of a
neutron star, this regime is achieved at temperatures $T\sim1\,\textrm{K}$, in which case the diffusion coefficient
given in Eq.~(\ref{diff_coefficient}) reduces to
\begin{eqnarray}\label{diff_coefficient_statonary}
 D&=&\frac{\kappa^2}{4}\Omega k_\textrm{B}T\sum_{jl\mu}\int\frac{d\vartheta(\alpha)}{2\omega_\alpha^2}
\frac{d\varpi}{2\varpi}
\left[\frac{|\phantom{}_1\Gamma_{l\mu}^{(j)}(\alpha,\varpi)|^2}{(\omega_\alpha+\varpi)^2+\Omega^2}\right.\nonumber\\
&&\left.+\frac{|\phantom{}_2\Gamma_{l\mu}^{(j)}(\alpha,\varpi)|^2}{(\omega_\alpha-\varpi)^2+\Omega^2}\right].
\end{eqnarray}
Then, by factorizing all the dimensional terms above, $D$ can be simply written as
\begin{equation}\label{diffusion_coefficient}
 D=8\pi\Delta\pp{\frac{\ell_\textrm{P}}{\mathsf{R}}}^2\pp{\frac{k_\textrm{B}T}{\Omega}}\Omega^2.
\end{equation}
In Eq.~(\ref{diffusion_coefficient}), $\Delta$ is a dimensionless quantity whose precise value will not be relevant to
estimate the decoherence time scale, although it is expected to be of order unity.

\subsection{Estimation of the decoherence time scale and the width of the pointer states}
\label{subsec:time_scale}

Here, we are interested only in the decoherence effects produced by the localization term in
Eq.~(\ref{long_time_regime_master_eq}), the master equation describing the unstable mode in its long-time regime. As
already mentioned --- see discussion below Eq.~(\ref{model_wf_eq}) --- the other terms appearing in that master
equation are essentially of two types: either they also produce decoherence but are not able to localize the
state in the amplitude and momentum representations or they are responsible for damping effects. While the former only
reinforce the consequences of the localization term in Eq.~(\ref{long_time_regime_master_eq}), the latter are not
important when the coupling with the environment is weak. Therefore, in order to estimate the decoherence rate in the
long-time regime, one can drop these terms and consider the following equations:
\begin{eqnarray}
 \del_t\varrho_\textrm{red}&=&-i\cc{\frac{1}{2}(-\del_q^2+\del_{q^\prime}^2)-\frac{\Omega^2}{2}(q^2-q^{\prime2})}
\varrho_\textrm{red}\nonumber\\
&&-D(q-q^\prime)^2\varrho_\textrm{red}
\end{eqnarray}
and
\begin{equation}\label{lindblad_wf_eq}
 \del_t W_\textrm{red}=\{H(q,p),W_\textrm{red}\}+D\del_p^2W_\textrm{red},
\end{equation}
with $H(q,p)$ given in Eq.~(\ref{class_hamiltonian}). 

A possible route to estimate the decoherence time scale is to investigate the temporal behavior of the Wigner function
sign. As mentioned earlier, the Wigner function is not positive in general. Nevertheless, decoherence should suppress
any negative region in the course of time. Indeed, it was shown in Ref.~\cite{dk_jpa_2002} that for a nonrelativistic
free quantum particle, the presence of a localization term in position in the master equation makes the corresponding
Wigner function positive after a certain time $t_\textrm{d}$, regardless of the initial state of the system. This result
was extended in Ref.~\cite{boa_pre_2004} for systems defined by quadratic Hamiltonians and with more general couplings
with the environment.

Following Refs.~\cite{dk_jpa_2002,boa_pre_2004}, the time $t_\textrm{d}$ after which the Wigner function
$W_\textrm{red}$ becomes positive is the solution of the equation
\begin{equation}\label{positivity_condition}
 \textrm{det}[\mathbf{M}(-t_\textrm{d})]=\frac{1}{4},
\end{equation}
where, in the case of our Eq.~(\ref{lindblad_wf_eq}),
\begin{equation}\label{det_M}
 \textrm{det}[\mathbf{M}(-t)]=\frac{D^2}{\Omega^4}\left[\frac{\cosh2\Omega t-1}{2}
-(\Omega t)^2\right].
\end{equation}
The feebleness of the gravitational interaction implies in our case that $D\ll\Omega^2$ and, thus, from
Eqs.~(\ref{positivity_condition}) and (\ref{det_M}) one obtains
\begin{equation*}
 \frac{2D^2}{\Omega^4}\cosh2\Omega t_\textrm{d}\approx1.
\end{equation*}
Then, by using Eq.~(\ref{diffusion_coefficient}), one concludes that
\begin{equation}\label{decoherence_time_scale}
 t_\textrm{d}\sim\frac{1}{\Omega}\ln\cc{\frac{1}{8\pi\Delta}\pp{\frac{\mathsf{R}}{\ell_\textrm{P}}}^2
\frac{\Omega}{k_\textrm{B}T}},
\end{equation}
which gives us an estimate for the decoherence time scale. The logarithm in Eq.~(\ref{decoherence_time_scale})
shows that the decoherence process depends weakly on the magnitude of the interaction with the environment, codified in
the diffusion coefficient --- see Eq.~(\ref{diffusion_coefficient}). This logarithmic dependence results from the
combination of the squeezing caused by the time evolution and the weak coupling between the unstable mode and its
environment. In particular, $t_\textrm{d}$ does not depend much on the value of $\Delta$. For the case of a neutron
star ($\mathsf{R}\sim10\, \textrm{km}$), if one assumes $\Delta\sim1$ and a cosmic gravitational wave background with
temperature $T\sim1\,\textrm{K}$, then $t_\textrm{d}\sim 160\times \Omega^{-1}\sim 160\times \mathsf{R}$, which is of
the order of the backreaction time scale $t_\textrm{br}\sim10^{-3}\, \textrm{s}$, when the vacuum and ordinary star
energy densities rival each other. We note that $\Omega t_\textrm{d}\gg1$, complying with the long-time approximation.

The investigation we have undertaken so far suggests that by the time backreaction becomes important, both the
appearance of classical correlations and decoherence have been effective to turn the unstable sector of the initial
vacuum state into a classically correlated statistical mixture of localized states in the amplitude and momentum
representations. The form of these pointer states depends on both the internal dynamics of the open quantum system and
its interaction with the environment, and it has been derived in just a few examples \cite{zhp_prl_1994,dk_prl_2000}.
However, within the simplifications made in this section, it is possible to estimate the width of these pointer states.
To do so, we shall again make use of the variables $u$ and $v$ defined in Eq.~(\ref{uv_variables}) to cast
Eq.~(\ref{lindblad_wf_eq}) as
\begin{eqnarray}\label{uv_lindblad_wf_eq}
 \del_t W_\textrm{red}&=&\Omega(u\del_u-v\del_v)W_\textrm{red}\nonumber\\
&&+\frac{D}{4}(\del_u^2+\del_v^2+2\del_{uv})W_\textrm{red}.
\end{eqnarray}
By assuming the weak-coupling limit, the state is squeezed along the $u$ direction and stretched along the $v$
direction, as concluded in Sec.\ \ref{sec:free_evolution} for free fields. Consequently, in the long-time regime, the
$u$ and $v$ derivatives grow and fade exponentially, respectively, and Eq.~(\ref{uv_lindblad_wf_eq}) can be cast as
\begin{equation}\label{lindblad_wf_eq_long_time_regime}
 \del_t W_\textrm{red}=\Omega(u\del_u-v\del_v)W_\textrm{red}+\Omega\sigma^2\del_u^2W_\textrm{red},
\end{equation}
with $\sigma^2\equiv D/4\Omega$.

Following the analysis of Ref.~\cite{zp_prl_1994}, a general solution of Eq.~(\ref{lindblad_wf_eq_long_time_regime})
can be expanded as
\begin{equation*}
 W_\textrm{red}=\sum_{\substack{m\ge0\\n\ge1}}^{+\infty}a_{mn}e^{-\Omega(m+n)t}v^me^{-\frac{u^2}{2\sigma^2}}H_{n-1}
\pp{\frac{u}{\sqrt{2}\sigma}},
\end{equation*}
where $H_n(x)$ denotes the $n$th Hermite polynomial. In the long-time regime, the sum above is dominated by the term
with $m=0$ and $n=1$ and is given approximately by
\begin{equation}\label{wf_long_time_regime}
 W_\textrm{red}(t,q,p)\approx\frac{e^{-\Omega t}}{\sqrt{2\pi\sigma^2}}e^{-\frac{u^2}{2\sigma^2}}
\int_{-\infty}^{+\infty}{du^\prime \tilde{W}_\textrm{red}(0,u^\prime,ve^{-\Omega t})},
\end{equation}
with $\tilde{W}_\textrm{red}(t,u,v)\equiv W_\textrm{red}[t,(v-u)/\Omega,v+u]$. Hence, when $\Omega t\gg1$ the effect of
decoherence on the Wigner function in the weak-coupling limit is to make $W_\textrm{red}$ approach a Gaussian in the
$u$ direction with width $\sigma$. This results from the competition between the free evolution, which tends to squeeze
the state in the $u$ direction, and the diffusive term in Eq.~(\ref{lindblad_wf_eq_long_time_regime}). Then, the width
of the density matrix in the $q$ representation can be obtained from Eq.~(\ref{wf_long_time_regime}) if we note that
\begin{eqnarray}
 \varrho_\textrm{red}(t,q,q^\prime)&=&\int_{-\infty}^{+\infty}{dpe^{ip(q-q^\prime)}W_\textrm{red}[t,(q+q^\prime)/2,p]}
\nonumber\\
&\approx&2e^{-2\sigma^2(q-q^\prime)^2-i\Omega(q^2-q^{\prime2})/2}e^{-\Omega t}\nonumber\\
&&\times\int_{-\infty}^{+\infty}{du^\prime\tilde{W}_\textrm{red}(0,u^\prime,0)}.
\end{eqnarray}
Thus, asymptotically, the initially pure density matrix becomes a statistical mixture of localized states with width
$(2\sigma)^{-1}=\sqrt{\Omega/D}$. By using Eq.~(\ref{diffusion_coefficient}) with $\Delta\sim1$, we obtain
\begin{equation}
(2\sigma)^{-1} \sim \frac{1}{\sqrt{k_\textrm{B}T}}\frac{\mathsf{R}}{\ell_\textrm{P}}
\end{equation}
in the case of our model. 

In the classical regime, one would like to regard each pointer state peaked at some amplitude and conjugate momentum as
a point in the phase space. This is possible only if the background spacetime is insensitive to quantum fluctuations
present in these states, i.e., if the pointer states are narrow enough. We can estimate how narrow these states are
through the energy-momentum tensor operator associated with the unstable mode. Thus, let us consider the contribution
from the quantum fluctuations to the expectation value of this operator in some pointer state. Taking, for instance, the
energy density, the contribution from a localized state with width $\sigma$ is $\rho_\textrm{qf} \sim
\sigma^{-2}\Omega^2/\mathsf{R}^3$. Then, the ratio between this contribution and the energy density $\rho$ of the
relativistic star curving the background is
\begin{equation*}
 \frac{\rho_\textrm{qf}}{\rho}\sim\frac{\Omega}{k_\textrm{B}T}.
\end{equation*}
Assuming $T\sim 1\, \textrm{K}$ and $\mathsf{R}\sim 10^4\, \textrm{m}$ --- the typical radius of a neutron star ---
the ratio above is of order $10^{-7}$. This last result shows that the pointer states of the unstable mode are narrow
enough in the long-time regime to be approximated by classical states.

In conclusion, by the time backreaction becomes ineluctable, the (unstable sector of the) initially pure vacuum
state has evolved into a statistical mixture of localized states in amplitude and momentum representations. The exact
form of these states and of the statistical weights can be calculated, in principle, from Eq.~(\ref{master_eq}). Since
the pointer states are narrow enough, one can regard these weights as a statistical distribution over the unstable mode
classical phase space, providing the initial conditions for the classical general-relativistic equations at the onset
of the backreaction.

We remark that the calculations assuming a graviton environment with temperature $T=0$ lead to the same master equation
as in the case with $T\neq 0$, namely, Eq.~(\ref{master_eq}). For a graviton environment in its vacuum state, we have
found that the decoherence time scale is of the same order as in Eq.~(\ref{decoherence_time_scale}), assuming $T\sim1\,
\textrm{K}$. As for the width of the pointer states, however, we have obtained $\rho_\textrm{qf}/\rho\sim1$. Hence, even
though decoherence diagonalizes the density matrix when $T=0$, it is not sufficiently effective to produce pointer
states which are narrow enough to be well approximated by points in the unstable-mode classical phase space.


\section{Conclusions and final remarks} 
\label{sec:final_remarks}

In conclusion, after the scalar field instability is triggered by the background spacetime, the interaction of the
field with gravity forces its quantum fluctuations to behave classically in a time scale of the same order of the one
set by backreaction. During this time, the appearance of classical correlations and decoherence are effective
enough to turn the vacuum state of the quantum field into a classically correlated statistical mixture of localized
states in the amplitude and momentum representations of the unstable mode. 

Then, we have argued how the gravity-induced vacuum dominance effect \cite{lv_prl_2010,lmv_prl_2010} gives rise to
classical initial conditions for the general-relativistic equations. This leads us to Ref.~\cite{pcbrs_prd_2011}, where
the authors discussed the possible final states for the instability in the spacetime of a relativistic star based on a
classical analysis. There it was shown that, at least for negative values of the nonminimally coupling parameter $\xi$,
the system can be stabilized by the presence of a nonnull, static scalar field profile. As for positive values of this
parameter, the final state remains an open issue. The appearance of a nontrivial classical field in the spacetime of
dense enough relativistic stars is analogous to the spontaneous magnetization of ferromagnets below the Curie
temperature and is known in the literature as ``spontaneous scalarization'' \cite{def_prl_1993,def_prd_1996}.
Typically, this phenomenon changes the gravitational mass of the star by a few percent and may have important
consequences for astrophysics \cite{e_bjp_2013}. Although the previous scalarization analyses do not consider quantum
mechanics to fix the initial conditions, the fact that their results seem to be robust with respect to the initial
conditions choice suggests that their conclusions should be preserved even when the instability is triggered by quantum
fluctuations. For more on the relation between the instability and the scalarization process, see, e.g.,
Refs.~\cite{mmv_prd_2014a,h_ptp_1997,n_prd_1998,rdans_prd_2012}.

The interest in scalar fields nonminimally coupled to gravity relies on the fact that most matter in the Universe cannot
be accommodated within the standard model of particle physics. Nonminimally coupled scalar fields have not been ruled
out by either astrophysical or cosmological observations so far, even though it is possible to put constraints on the
values of the nonminimally coupling parameter --- see, e.g., Ref.~\cite{freire_etal_mnras_2012}.

\acknowledgments

A.~L.\ was partially supported by Conselho Nacional de Desenvolvimento Cient\'\i fico e Tecnol\'ogico (CNPq) under
Grant No.~610020/2009-9 and S\~ao Paulo Research Foundation (FAPESP) under Grant No.~08/57856-6 through Instituto
Nacional de Ci\^encia e Tecnologia de Informa\c c\~ao Qu\^antica (INCT- IQ). W.~L.\ was supported by FAPESP under Grant
No.~2012/00737-0. G.~M.\ and D.~V.\ acknowledge partial support from FAPESP under Grants No.~2007/55449-1 and
No.~2013/12165-4, respectively. G.~M.\ also acknowledges CNPq for partial support.

\appendix
\section{Expressions for the tensors $U_{abcd}$, $V_{abcde}$, and $W_{abcdef}$}
\label{appendix}

In this appendix we present the expressions for the tensors $U_{abcd}$, $V_{abcde}$, and $W_{abcdef}$ appearing in Sec.\
\ref{sec:decoherence}. By defining the tensor $I_{abcd}\equiv\frac{1}{2}(g_{ac}g_{bd}+g_{ad}g_{bc})$, one has
\begin{widetext}
\begin{eqnarray}\label{U_tensor}
 U_{abcd}&\equiv&
-(1-2\xi)(g_{ad}\nabla_b\phi\nabla_c\phi+g_{bc}\nabla_a\phi\nabla_d\phi)
+\frac{1-4\xi}{2}(g_{ab}\nabla_c\phi\nabla_d\phi+g_{cd}\nabla_a\phi\nabla_b\phi)\nonumber\\
&&+\frac{1-4\xi}{2}\pp{I_{abcd}-\frac{1}{2}g_{ab}g_{cd}}\nabla_e\phi\nabla^e\phi
+\pp{\frac{1}{2}I_{abcd}-\frac{1}{4}g_{ab}g_{cd}}(m^2+\xi R)\phi^2\nonumber\\
&&-\xi\left[(2I_{abcd}-g_{ab}g_{cd})\phi\nabla_e\nabla^e\phi
+\left(g_{ad}R_{bc}+g_{bc}R_{ad}-\frac{1}{2}g_{ab}R_{cd}-\frac{1}{2}g_{cd}R_{ab}
\right)\right.\nonumber\\
&&\left.\phantom{\pp{\frac{1}{1}}}+2(g_{ab}\phi\nabla_c\nabla_d\phi+g_{cd}\phi\nabla_a\nabla_b\phi
-g_{ad}\phi\nabla_b\nabla_c\phi-g_{bc}\phi\nabla_a\nabla_d\phi)\right],
\end{eqnarray}
\begin{equation}\label{V_tensor}
 V_{abcde}\equiv-g_{ab}(g_{cd}\phi\nabla_e\phi+g_{ce}\phi\nabla_d\phi),
\end{equation}
and
\begin{equation}\label{W_tensor}
 W_{abcdef}\equiv\frac{1}{2}\cc{g_{ad}I_{bcef}+2g_{ef}I_{adbc}-g_{ad}g_{bc}g_{ef}
-\frac{1}{2}(g_{ae}g_{bd}g_{cf}+g_{ae}g_{cd}g_{bf}+g_{af}g_{bd}g_{ce}+g_{af}g_{cd}g_{be})}
\phi^2.
\end{equation}
\end{widetext}



\begin{thebibliography}{}
\bibitem{footnote_2} We note that the second term in Eq.~(\ref{int_second_order}) can be absorbed into the first one
if one uses the de Donder gauge. The gauge choice in this paper is discussed below Eq.~(\ref{int_action}).
\end{thebibliography}
\end{document}